\documentclass[fleqn,10pt]{wlscirep}
\usepackage[utf8]{inputenc}
\usepackage[T1]{fontenc}
\usepackage[super]{nth}
\usepackage[noabbrev, capitalise]{cleveref}
\usepackage{url}
\usepackage[labelfont=bf, font+=footnotesize]{subcaption}
\title{The Paradox of Talent: how Chance affects Success in Tennis Tournaments}

\author[1]{Chiara Zappal\`a}
\author[2]{Alessio Emanuele Biondo}
\author[1]{Alessandro Pluchino}
\author[1,3,*]{Andrea Rapisarda}
\affil[1]{Department of Physics and Astronomy, University of Catania and INFN sezione di Catania, Catania, 95123, Italy}
\affil[2]{Department of Economics and Business, University of Catania, Catania, 95129, Italy}
\affil[3]{Complexity Science Hub Vienna, Vienna, 1080, Austria}
\affil[*]{andrea.rapisarda@unict.it}

\begin{abstract}

Individual sports competitions provide a natural setting for examining the relative importance of talent and luck/chance in achieving success.
The belief that success is primarily due to individual abilities and hard work rather than external factors is particularly strong in this context.
Thus, individual talent is regarded as the most important – if not the only – component in ensuring a successful career for athletes.
In this study, we test this belief using tennis as a case study, due to its popularity and competition structure in direct-elimination tournaments. 
Our dataset covers the last decade before Covid-19 pandemics (2010-2019) of main international events in the ATP circuit and consists of tourney results and annual rankings for professional male players.
After a preliminary data analysis, we introduce an agent-based model able to accurately simulate the tennis players' dynamics along several seasons. We show that, once calibrated on the dataset, the model is able to reproduce the main stylized facts observed in real data, including the results of single tournaments and the development of players' careers in the ATP community.
The strength of our approach lies in its simplicity: it requires only one free parameter $a$ to determine the importance of talent in scoring every single point: $a = 1$ indicates the ideal scenario in which only talent matters, whereas $a = 0$ represents the opposite limit case, in which the outcome of each point is entirely due to chance. 
We find the best agreement between real data and simulation results when talent weights substantially less than luck, i.e. when $a$ is between $0.20$ and $0.30$. A further comparison between data and simulations, based on the analysis of the direct networks of all the matches, confirms the previous finding.  
\textit{A posteriori}, we notice that this surprisingly important role of chance in tennis tournaments is not an exception.
On the contrary, it can be explained by a more general paradoxical effect that characterizes highly competitive environments, particularly in individual sports.
In other words, \textit{when the difference in talent between top players is minimal, chance becomes determinant.}
Our findings highlight the impact of the -- too often underestimated -- external factors on athletes' performance in individual tournaments and their careers.
Our results  point out the unfairness of the ``winner-takes-all'' reward system that enhances the disparities between the first classified players and the others in major competitions.

\end{abstract}
\begin{document}

\flushbottom
\maketitle
%
%
\thispagestyle{empty}

\section{1. Introduction}
\label{sec:intro}

Understanding the origin of success is a challenging task for researchers in complex systems.
Science of Success has recently received a lot of attention because it concerns a huge variety of systems: e.g., paper citations in science \cite{Sinatra2016}, arts \cite{Fraiberger2018} and creative careers in general, show business \cite{Williams2019}, and so on.

Investigating what leads to success is inevitably correlated to the role that chance might play in achieving it.
We are reluctant to acknowledge the presence of luck in our achievements, as if it could reduce our legitimacy to them.
Instead, we adopt two tendencies to neglect the influence of random events: the hindsight bias and the \emph{post hoc} fallacy. 
The former describes the human predisposition to think that events are more predictable than they are \cite{Frank2016}.
The latter comes from the Latin ``\emph{post hoc, ergo propter hoc}'', which means ``after this, therefore because of this'' and expresses the propensity to attribute causality to two events just because one happened after the other \cite{mauboussin2012success}.
Those two ways of thinking explain why we systematically ignore or underestimate the contribution of chance.

Nevertheless, the influence of unpredictable factors does not necessarily reduce the importance of individual skills.
However, the evidence suggests that a combination of talent and effort does not guarantee success \cite{Frank2016,Pluchino2018,Pluchino2018a}.
Successful people must also be very lucky, other than talented and hard-working \cite{Frank2016}.

Despite the increasing amount of literature on this subject, there has been little attention on analysing success in sports.
In principle, sports should be the ideal systems to be examined, given their clear and objective measures of performance and well-defined rules in very controlled environments, other than their ability to attract widespread interest.
Yet, only a handful of sports, such as soccer \cite{Lago2010}, baseball \cite{Petersen2008}, and tennis \cite{Yucesoy2016}, have been studied, and chance has seldom been considered \cite{Sobkowicz2020}.
Even with our constant exposure to success stories about famous athletes and our intuitive sense that luck plays a role in becoming a top player, we still assume sports to be fundamentally meritocratic.

Those arguments motivated us to investigate the role of chance in sports, and in particular in individual sports, quantitatively.
Specifically, we are interested in individual sports where athletes face an opponent and their competitions are structured as knock-out tournaments.

In this study, we analyse tennis, which has a broad audience and provides easy access to large and detailed datasets about the main international tournaments, inspired by previous work on fencing \cite{Zappala2022}.
In tennis, as in fencing, one of the main goals for a player is to rise in the official ranking.
In an ideal talent-oriented view, one could expect a perfect correspondence between talent and ranking order.
Yet, athletes very close in talent might end up with totally different rewards for similar performances, particularly when their talent is quite high.
The selective mechanisms behind direct-elimination tournaments can systematically amplify little differences between players \cite{McGarry1997} since those kinds of competitions allow athletes to overcome many opponents at the same time, simply by defeating one of them in a given round.

In individual sports based on knock-out tournaments, small random fluctuations in performances might trigger a rich-get-richer phenomenon, generating a consistent gap in the ranking between individuals with comparable skills.
This process is accelerated by the specific scaling of ranking points, gained by players after a competition.
Those scales, usually chosen \emph{ad hoc}, often follow a nonlinear trend which is hard to justify, given that many features included in the word ``talent'' (intelligence, physical characteristics, skills, mental strength, endurance and abilities in general) are typically normally distributed in the population \cite{Stewart1983,Pluchino2018}.
This promotes an unfair ``winner-takes-all'' rationale, thus emphasizing disparities among the athletes and increasing the influence of random events.

In this work, we untangle the relative effects of random events and individual abilities by analysing the dynamics of a community of athletes in the ATP (Association of Tennis Professionals) circuit.
We use an agent-based approach to assess the impact of randomness in tennis tournaments by comparing the results with real data.
Given that «complexity can grow out of simplicity» \cite{Buchanan2002}, we rely on a simple linear model, already adopted in previous works \cite{Pluchino2018,Pluchino2018a,Sobkowicz2020,Zappala2022}. 
In detail, we assume a certain weight, expressed by a unique constant parameter $a$, to quantify the relative role of talent and chance for each point in the match.

We show that our model can reproduce not only athletes' performance (in terms of match victories/defeats) but also the overall dynamics of their careers over several years. 
Moreover, within a small range of (unexpectedly low) values for the parameter $a$, our model can provide a strongly nonlinear ranking whose statistical features are in good agreement with those of the real one.
Additionally, based on the real and the simulated players' ranking, we build and compare the directed {\it loser-to-winner} networks of the whole sequence of considered matches.
Through those networks, we highlight the existence of a paradoxical mechanism in highly competitive environments: as skill improves, performance becomes more consistent, and therefore chance becomes more relevant \cite{mauboussin2012success}.

Our results advocate a rethinking of the meritocratic paradigm to align with recent scientific discoveries about the role of chance in individual sports.

The rest of the paper is organized as follows: in Section 2 we present our agent-based model, the dataset adopted for the comparison with real data and the calibration of the model with these data; in section 3 we present our main results; finally, in Section 4, we close the paper with some conclusive remarks. 

\section{2. Dataset Analysis and Model Calibration}
\label{sec:results}

Some sports, among which tennis, allow players to win a match even after winning less points than their opponent.
Such evidence suggests that sport rules might create an inherent advantage during the game, leading to an unbalanced and unfair result.
This is true especially in sports based on tournaments, where the winner overcomes not only his/her direct opponent, but also all the other losers of that match round.
That advantage is reflected on the amount of points gained after a competition, which scales non linearly as a function of the final ranking.

In order to investigate the aspects that create disparities in ATP ranking and to estimate the weight of individual abilities on these disparities, as opposed to random external influences, we built and agent-based model able to reproduce the dynamics of several competitive seasons in professional male tennis, starting from a few assumptions at the level of single points.

Tennis is a racket sport that is usually played individually against a single opponent, in a rectangular court with a net across the center \cite{tennisrules}.
The objective is to hit the ball over the net so that it lands inside the court's margins and cannot be returned by their opponent.
When that happens, players score a point \cite{tennisrules}.
One need at least four points to win a game and six games to win a set, in both cases with a margin of two in case of equal score.
A match is usually played as best of three or five sets (Slam tournament) \cite{tennisrules}.
If a game score of 6-6 is reached, players must play a tie-break game in order to decide who wins the set.
In a tie-break game, a player must reach 7 points with a two point advantage to win \cite{tennisrules}.

Since every point weights differently in the economy of a match, and one cannot simply score more points than their opponent to win, the influence of external factors can either be marginal or crucial.
In this framework, the interchange between serve and return become relevant and must be taken into account when modelling a single point in tennis.

\begin{figure}[h!]
\begin{subfigure}{.5\textwidth}
\centering
\includegraphics[width=\linewidth]{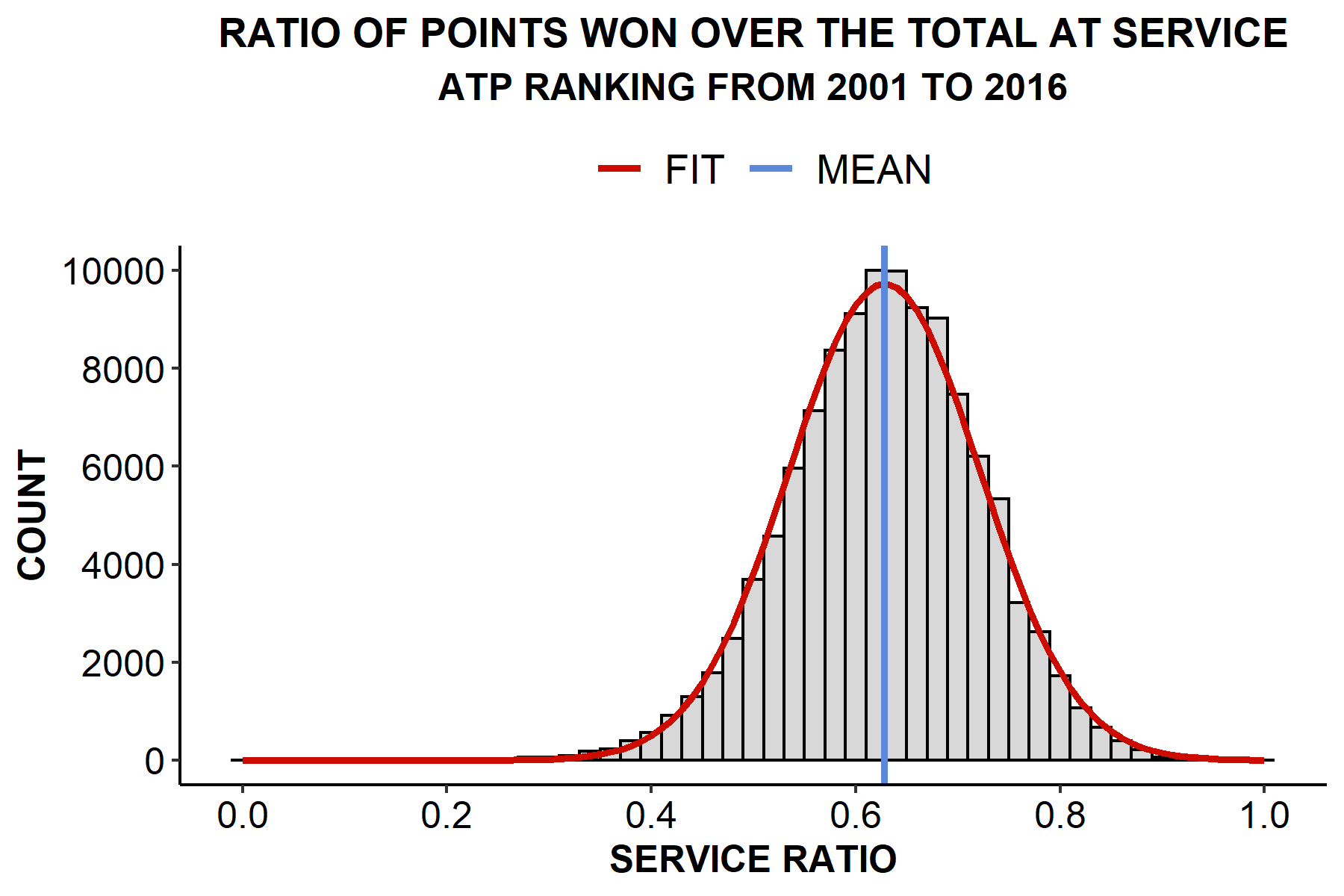}
\subcaption{Probability of winning a point at service.}
\label{fig:serv}
\end{subfigure}%
\begin{subfigure}{.5\textwidth}
\centering
\includegraphics[width=\linewidth]{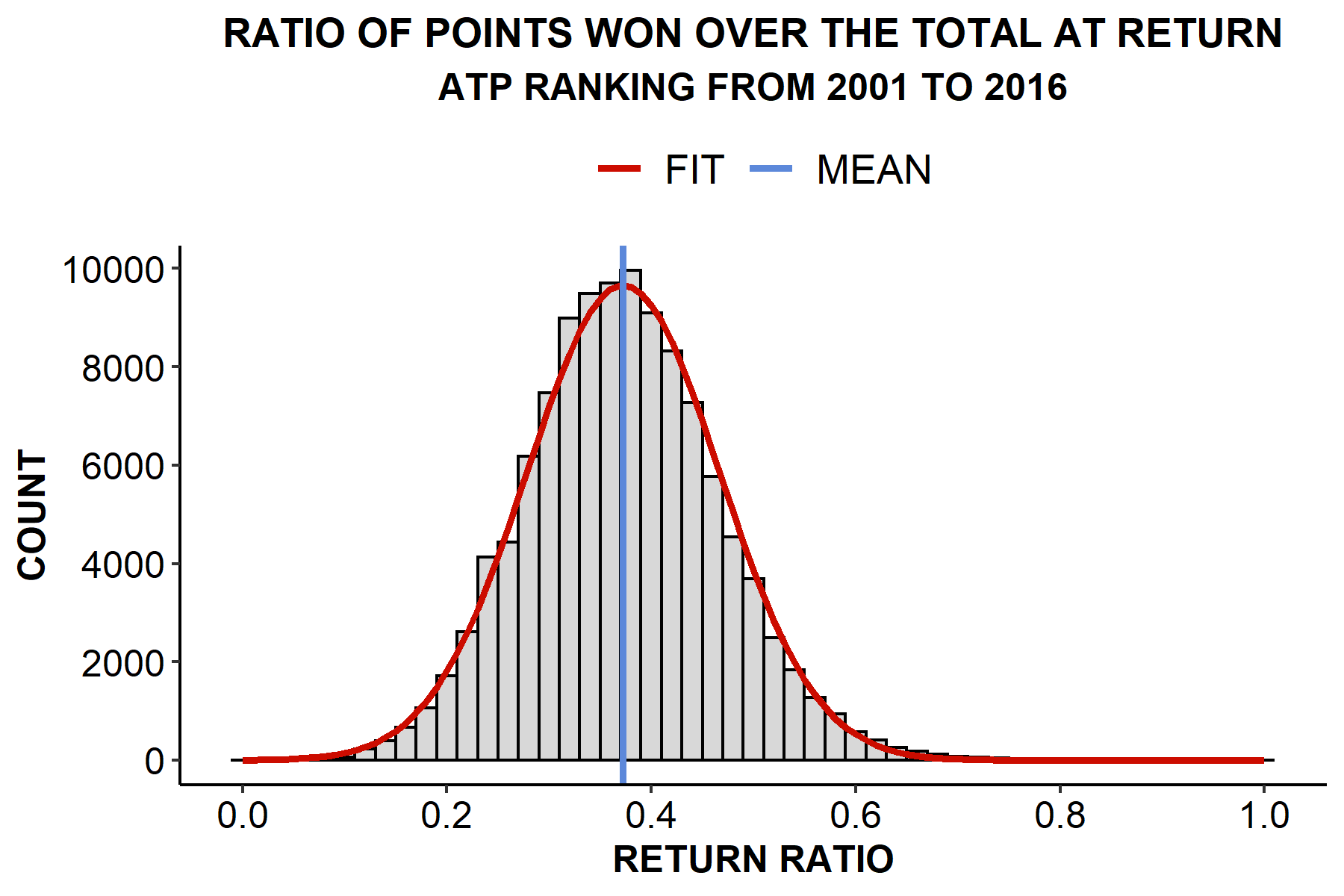}
\subcaption{Probability of winning a point at return.}
\label{fig:ret}
\end{subfigure}%
\caption{Gaussian distributions of points won at service (left) and return (right) over the total extracted from the ATP dataset, see text.}
\label{fig:serv_ret}
\end{figure}

\subsection{2.1 Modelling and calibration of serve and return points}
\label{subsec:serv_ret}

During a single match of a tournament, organized in sets and games, random external factors act at the local level of single points won or lost, but they can have a different impact conditional to the rules of tennis.
Moreover, the natural asymmetry of this sport, due to players switching between service and return, adds some noise to matches, affecting macroscopically both the outcome of tournaments and the development of players' careers.

To take into account these peculiar features at the level of each single point, in our model, we considered two distinct performances, $P^{S}$ and $P^{R}$ (with $P^{S}, P^{R} \in \Re$), for players scoring at service and at return, respectively.
These performances emerge from a combination of both players' talent and chance, weighted by a control parameter $a \in [0,1]$, and are defined as follows:
\begin{equation}
\label{eq:serv}
\\ P_{i}^{S} = aT_{i} + \left( 1 - a \right) L_{i}^{S} \\
P_{j}^{R} = aT_{j} + \left( 1 - a \right) L_{j}^{R}
\end{equation}
In these equations $i = 1, 2$ and $j = 2, 1$ are the players' indexes, fixed for a given match; $T_k$ ($k = i,j$) is the talent of the player, which captures all the individual skills of athletes (height, intelligence, ability, training, etc.); $L_k^S$ and $L_k^R$ ($k = i,j$) are the values of the chance of the agents, which randomly affect the possibility of winning a point for the same athletes when they are at service or at return, respectively.

Imagine that player $1$ is at service and, consequently, player $2$ is at return: if $P_{1}^{S} > P_{2}^{R}$, $P_{1}$ gets the point; vice-versa, if $P_{2}^{R} > P_{1}^{S}$, the point is assigned to $P_{2}$. Once they move to the next game of the set, players $1$ and $2$ reverse their previous role, so that analogous conditions apply to get a point: if $P_{2}^{S} > P_{1}^{R}$, $P_{2}$ scores; conversely, if $P_{1}^{R} > P_{2}^{S}$, $P_{1}$ scores.

Inspired by previous studies \cite{Pluchino2018,Sobkowicz2020,Pluchino2018a,Pluchino2020,Zappala2022}, $T\in(0,1)$  is extracted from a Gaussian distribution with $\mu = 0.6$ and $\sigma = 0.1$ at the beginning of every simulation run, thus it is fixed for every player.
$L^{S}\in(0,1)$ and $L^{R}\in(0,1)$ are also extracted from Gaussian distributions, with different means, $\mu^{S} = 0.6$ and $\mu^{R} = 0.4$, and the same $\sigma^{S,R} = 0.25$; but, unlike $T$, they change for every point. We put a constraint on $L^{S}$ and $L^{R}$, to consider the possibility of aces or return aces. In particular, if the $i$-th extracted value $L_{i}^{S}$ is greater than one, the point is assigned to the server $i$, simulating a winning shot at serve (ace); on the other hand, if the $j$-th extracted value $L_{j}^{R}$ is greater than one, the point goes automatically to the receiver $j$, as in the case of a return ace, e.g. a winning shot of the receiver returning a serve.
Notice also that the parameter $a$, weighting the relative contribution of talent as opposed to chance in  \cref{eq:serv}, remains the only free global parameter of the model.
Of course, $a = 1$ would indicate the ideal scenario where only talent matters while $a = 0$ the opposite, limit case, in which the outcome of each point is completely due to external random forces. 

Values of $\mu^{R}$, $\mu^{S}$ and $\sigma^{S,R}$ have been chosen after an accurate calibration in order to model the actual advantage of the serving player.
The calibration and testing of the model is based on data collected from several sources \cite{ATP,ATPdatahub,ATPgit}; we considered all available matches from $2010$ to $2019$ of the first thousand players in ATP ranking during the corresponding seasons, related only to main tournaments of the ATP Tour: Grand Slams, Masters 1000, ATP 500 and ATP 250.
For each year, we selected the top $1000$ players according to the last update of their ranking.
Correspondingly, in our model, we simulated a community of $1000$ agents, assuming they do not change over the years (this is a first approximation of the real case).
We also fixed the number of seasons $N_{S} = 15$ for each simulation run, discarding the first five years to ensure ranking reliability.
Any agent has a probability of attending a certain number of tournaments ($\leq N_{T}$) in a given season, conditional to the ranking order of that year.

\begin{figure}[h!]
\centering
\includegraphics[width=.51\textwidth]{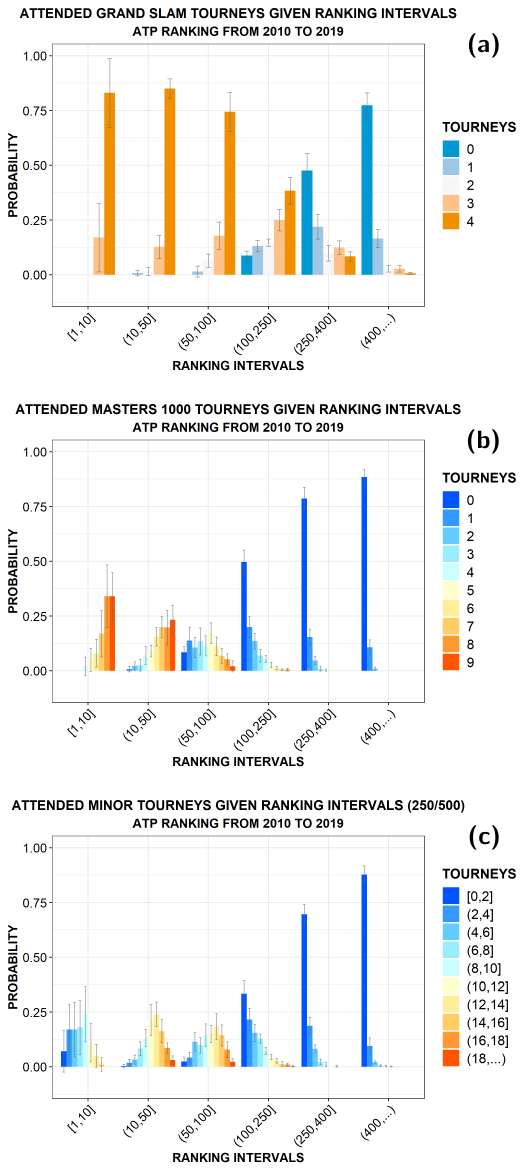}
\caption{Probabilities of attending different kind of tournaments extracted from our ATP dataset (panels (a), (b), (c) refer to Grand Slams, Masters 1000, ATP 500 and 250 respectively), as a function of ranking order.
Ranking positions are grouped in intervals to enhance visualization.}
\label{fig:bardata}
\end{figure}

In detail, we simulated $60000$ matches for $1000$ players, at best of 3 sets, with each point scored according to \cref{eq:serv}, testing several values for $\mu^{R}$, $\mu^{S}$ and $\sigma^{S,R}$ in order to reproduce the real probability distributions of points won as servers or receivers obtained from our dataset (see \cref{fig:serv_ret}), deriving from $57877$ matches of $2723$ players.
As expected, these distributions are not centered around $0.5$, showing the natural asymmetry --we could say, intrinsic-- in the tennis rules, where chances of winning a point at service are much higher than those at receiving. 

\subsection{2.2 Calibrating athletes' participation to ATP tournaments}
\label{subsec:calib}

With our agent-based model we aim to reproduce, within the NetLogo environment \cite{NetLogo}, the dynamics of the main tournaments of ATP Tour, which is the most relevant circuit for men's professional tennis players.
To do this, we simulate a community of $1000$ agents (athletes) playing for $N_{S} = 10$ seasons, each composed by a given number $N_{T}$ of tournaments.
Any agent has a different probability of attending a certain number $\leq N_{T}$ of events in a given season, conditional to his ranking order in that year.
As shown in the three panels of \cref{fig:bardata}, we calibrated these probabilities exploiting the information present in our dataset and related to the first $1000$ athletes in the ATP ranking who attended the main tournaments in seasons from $2010$ to $2019$ \cite{ATP,ATPdatahub,ATPgit}.
From data in panels (a) and (b) one may assume that the higher their ranking in a given season, the more events athletes played in that season.
In tennis, this is not generically correct, because of the very different weight characterizing distinct types of tourneys. 

\begin{figure}[t]
\centering
\includegraphics[width=\textwidth]{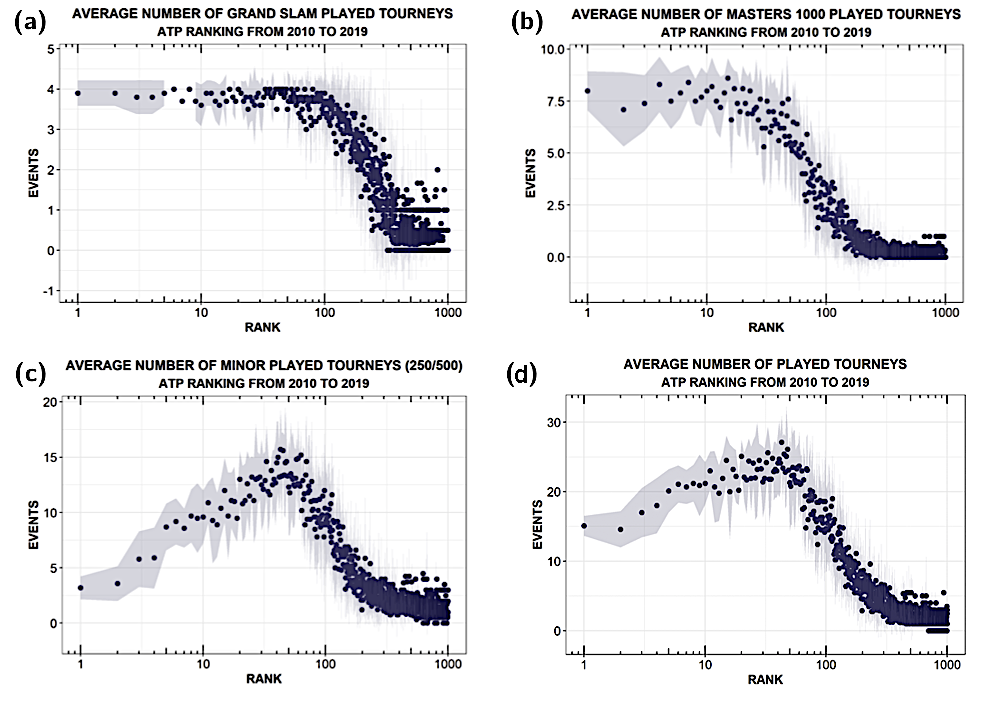}
\caption{Average number of played tournaments as a function of ranking placements in ATP tour extracted from our dataset.
Panels (a), (b), (c) refer to Grand Slams, Masters 1000, ATP 500 and 250 respectively
 Panels (d) refers to all the tournaments.}
\label{fig:ntour}
\end{figure}

In fact, looking at the average number of the attended competitions, reported in \cref{fig:ntour} as function of the ranking placement, and comparing this number calculated for all the tournaments (panel d) with the same one disaggregated per each tournament category (panels a, b and c), different patterns emerge: the top 10 ranked players tend to compete less than athletes in slightly lower positions (from 10 to 50), and have a comparable number of tourneys with respect to players in the range $\left[50, 100\right]$.

This might be explained by the large amount of ATP 500 and 250 (sometimes we refer to them as ``minor'' tourneys to be concise) in contrast with Grand Slams and Masters 1000, but also by ATP rules: only the best eighteen results are considered in ATP ranking \cite{ATP}, hence top 10 players, who have easier access to major events, tend to compete in the minimum amount of tournaments required; on the other hand, athletes reasonably close to the top 10 (up to the \nth{100} position), try to improve their placement by increasing the number of minor tourneys they can attend.
External factors could also affect athletes' participation, like prize money, funding availability, country of origin and of the tournament, injuries, and so on.
All those elements result in the non-monotonic trend displayed in \cref{fig:ntour}.

\begin{figure}[h]
\centering
\includegraphics[width=0.7\textwidth]{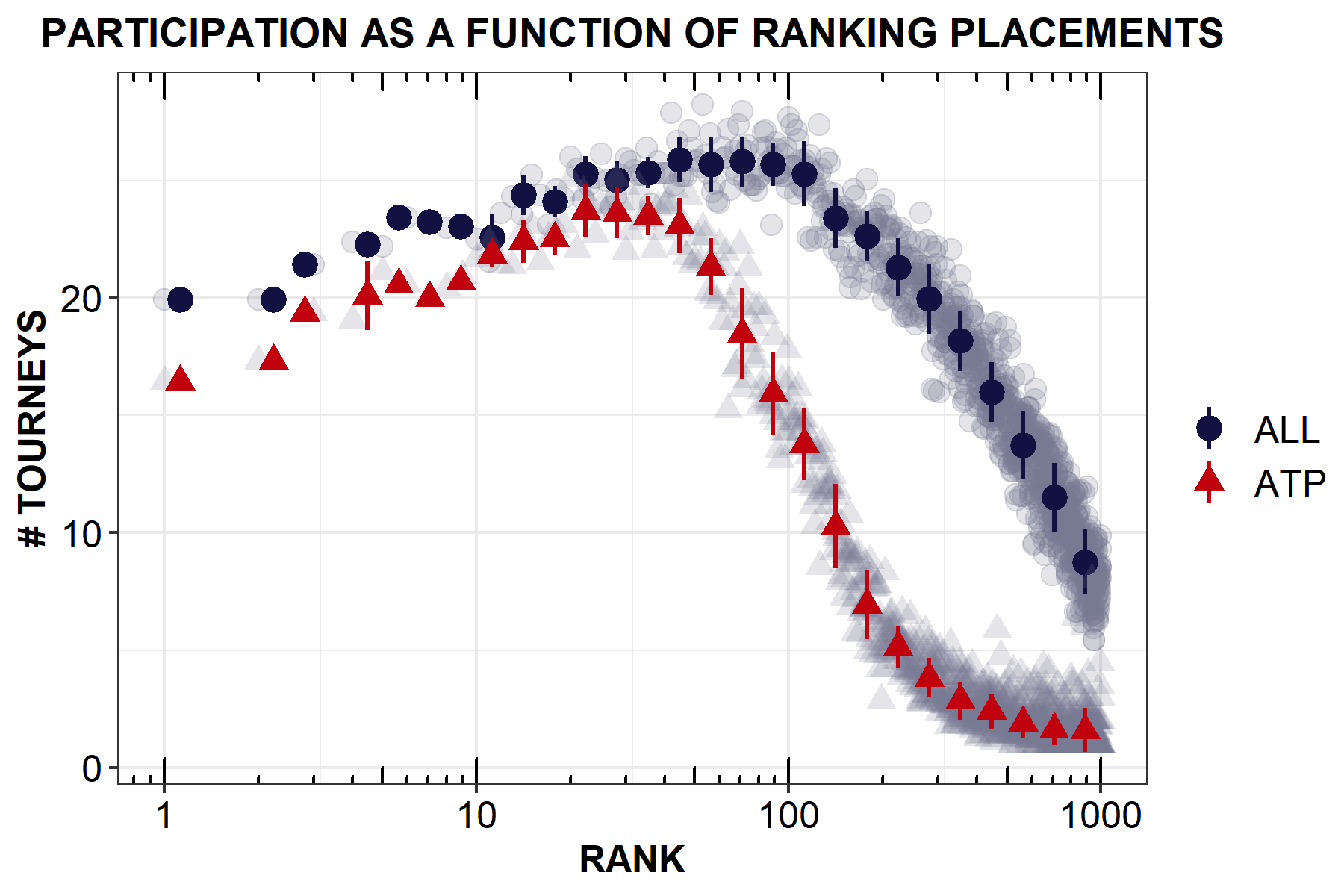}
\caption{Tourney attendance as a function of ATP ranking, considering all kind of competitions (blue dots) or just the main events (red triangles). Taken from our ATP dataset, see text}
\label{fig:ntour_comp}
\end{figure}

Since we restrict our dataset to the main tournaments of the ATP tour, one might ask whether they are enough to explain the evolution of ATP ranking in our dataset.
\cref{fig:ntour_comp} shows the difference between the total number of tournaments considered in the ranking (blue dots) and the number of ATP tournaments present in our dataset (red triangles), as a function of ranking placements.
There is a gap from the \nth{50} position, which widens for lower placements, meaning that ranking is shaped by other tourneys, not present in our dataset.
Given that, we expect our model can better capture the trend of the first fifty athletes, in terms of ranking points, while it might highlight other features of the system.

\subsection{2.3 Simulating ATP tournaments}
\label{subsec:ATP_tour}

\begin{figure}
\centering
\includegraphics[width=0.7\textwidth]{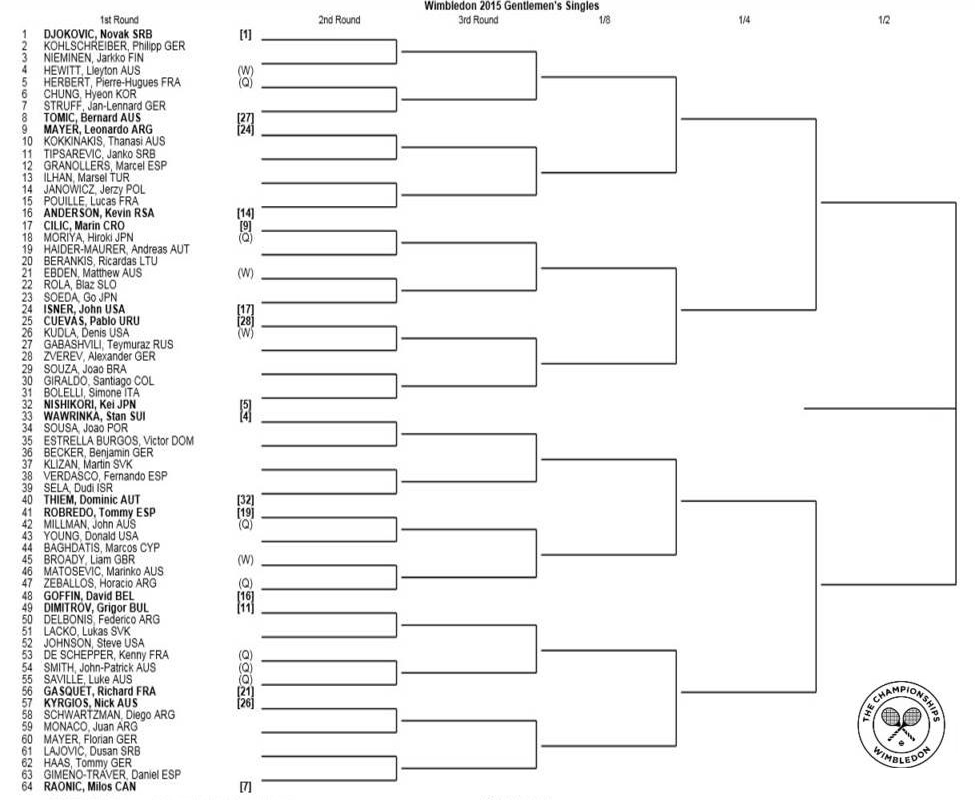}
\caption{Example of a main draw (upper part), from Wimbledon 2015 (source: \url{https://www.dotsport.it/}).
Seeds are highlighted in bold, placed in such a way they do not face each other until the \nth{3} round.
Wild cards and qualifiers are underlined by the symbols (W) and (Q) respectively.}
\label{fig:draw}
\end{figure}

As anticipated, in our model we only consider the most important tourneys in terms of ranking points, prizes and prestige: Grand Slams, Masters 1000, ATP 500 and ATP 250.
Tournament draws in these competitions are arranged in rounds that halve players at each step of the tournament. Each competition has two knock-out phases, thus two distinct draws: a qualifying one followed by the main one.
The qualifying draw allows those players who cannot directly access the main phase of a certain tournament to qualify for a place in the starting line-up of that tournament, after facing a preliminary set of rounds.
Once the qualifiers have been determined, the main draw can start.
Both draws are arranged in an analogous way, placing seeded players first, then all the others (see \cref{fig:draw} for a visual example).

The number $N$ of participants per tournament is fixed and related to its type. To compete in a given tournament, players are primarily selected according to their ranking placement, but there are also other criteria, which might leave space to randomness.
This is the case of wild cards, which are a small number of reserved spots that can be assigned to randomly selected players.
For example, wild cards may be exploited by local players who do not gain direct acceptance, or by players who are just outside the ranking range required to gain direct acceptance, or even for players whose ranking has dropped due to a long-term injury \cite{Shine2003}. 

There are other reserved places for the so-called seeded players we mentioned above.
Seeds are players whose position in a tournament is arranged based on their ranking in order to prevent them from encountering other seeds in the early rounds of the competition.
For a given tournament, there is a specified number of seeds, depending on the size of the draw.
For ATP tournaments, typically one out of four players are seeds.
The seeds are usually chosen and ranked by the tournament organizers \cite{Hedges1978}.

For the sake of simplicity, we refer to the official ATP rulebook \cite{ATP}, but we make some approximations to model the dynamics of several tennis tournaments.
In \cref{tab:tennis_numbers} we summarise the values we use in our agent-based model to arrange tournament draws, based on ATP rules.

\begin{table}[h]
\centering
\begin{tabular}{lcccc}
\hline
 & Grand Slam & Masters 1000 & ATP 500 & ATP 250 \\
\hline
\small{N-participants} & 128 & 128/64 & 32 & 32 \\
\small{N-seeds} & 32 & 32/16 & 8 & 8 \\
\small{N-qualif-participants} & 128 & 64/32 & 16 & 16 \\
\small{N-qualif-seeds} & 32 & 16/8 & 4 & 4 \\
\small{N-qualifiers} & 16 & 16/8 & 4 & 4 \\
\small{N-wild-cards} & 8 & 5/4 & 3 & 3 \\
\hline
\end{tabular}
\caption{Summary of the composition of tournaments we used in our model.}
\label{tab:tennis_numbers}
\end{table}

In our model we limit the number $N_{T}$ of competitions during each season to one tournament per week (except for Grand Slams which can last two weeks), so that there are $N_{T} = 52$ events in total.
In particular, we simulate four Grand Slams, nine Masters 1000, thirteen ATP 500 and twenty-six ATP 250.
This choice avoids general tourney overlaps during a simulated year, and is consistent with a real tennis season.
The only exception concerns ATP 250 tournaments, which should be around 36.
However, such an underestimation of ATP 250 contribution improves model interpretability, without decreasing its predictive power; moreover, it requires less computational resources.

We assume there are no byes in the first rounds, meaning that all the players must face an opponent.
Consequently, we increase the number of participants in Masters 1000 to 128 and 64, while they should be 96 and 56, in order to complete the draws.
It is worth noting that Masters 1000 are the only tournaments with two alternative lengths of their draws, whereas in every other case we consider only one possible draw size.

In \cref{tab:tennis_points} we show the specific scales of points for different types of competitions. We can see that they all decrease in a non-linear way.
Accumulating points let athletes rise in the official ATP ranking, but they are only valid for one season: any new result cancels out the corresponding one of the previous year, if present.

\begin{table}[h]
\centering
\begin{tabular}{lcccc}
\hline
 & Grand Slam & Masters 1000 & ATP 500 & ATP 250 \\
\hline
Winner & 2000 & 1000 & 500 & 250 \\
Final & 1200 & 600 & 300 & 150 \\
Semi-Final & 720 & 360 & 180 & 90 \\
Quarter-Final & 360 & 180 & 90 & 45 \\
Round-of-16 & 180 & 90 & 45 & 20 \\
Round-of-32 & 90 & 45 & 20 & 10 \\
Round-of-64 & 45 & 20 & - & - \\
Round-of-128 & 10 & $10^{*}$ & - & - \\
Qualif-\nth{1} & 25 & 20 & 20 & 10 \\
Qualif-\nth{2} & 16 & 10 & 10 & 5 \\ 
Qualif-\nth{3} & 8 & - & - & - \\
\hline
\end{tabular}
\caption{Allocation of points per tournament and round.}
\label{tab:tennis_points}
\end{table}

\begin{figure}[h!]
\centering
\includegraphics[width=0.9\textwidth]{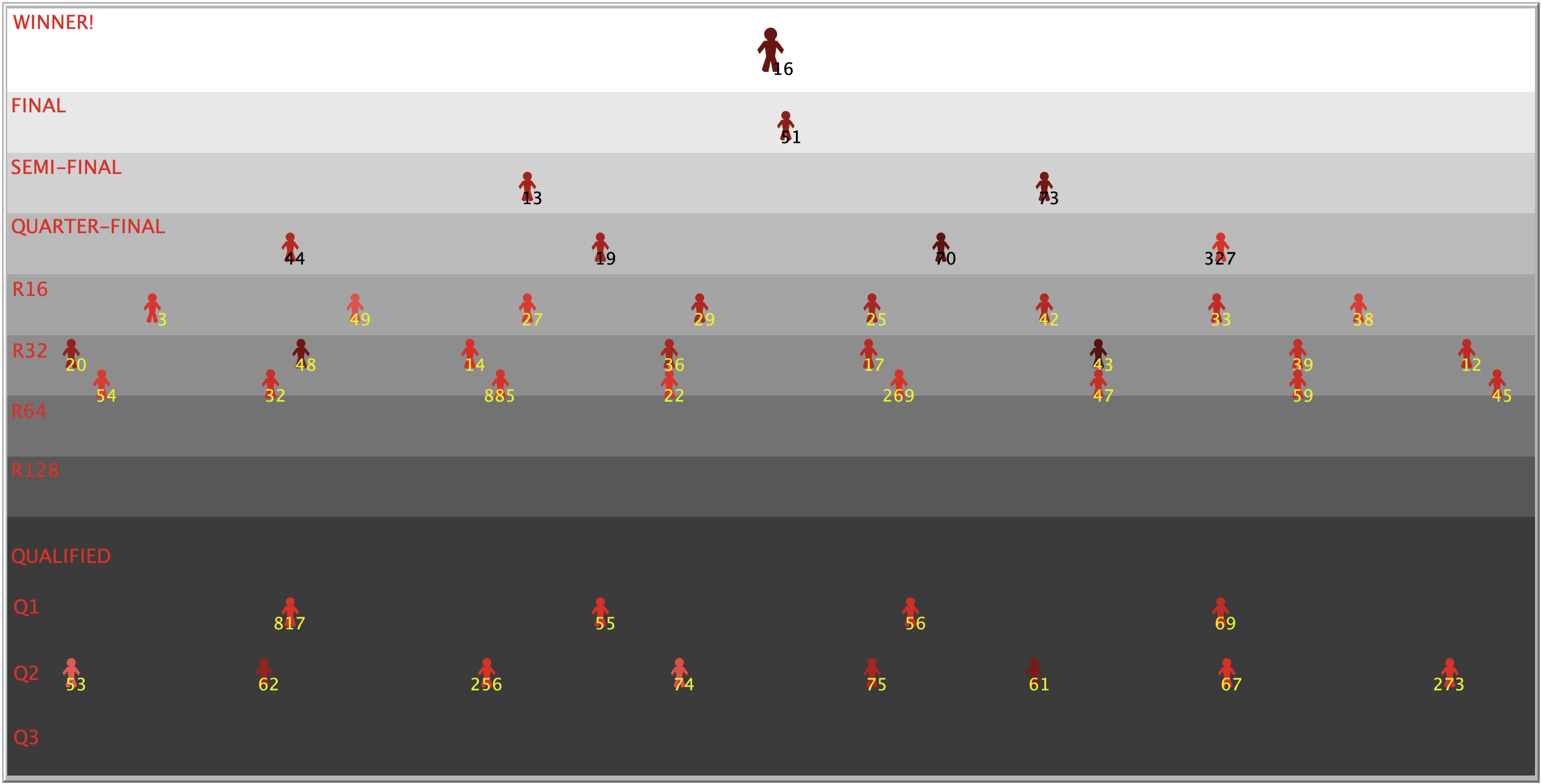}
\caption{An example of the output in the NetLogo interface at the end of a single tournament simulated with our agent-based model. At the bottom, the agents/athletes who did not pass the qualification phase are visible; at the higher levels the agents who passed the various rounds are also visualized until, at the top level, the winner is highlighted with a greater size. The color intensity of the agents is proportional to their talent, while the numbers in the labels represent the ranking of the agents at the beginning of the tournament. In this example, the $16$-th athlete of the ranking won the tournament.}
\label{fig:model-draw}
\end{figure}

Every run of our simulations starts with a random allocation of the $1000$ players in an initial ranking, which is updated as soon as agents attend events and earn points, following the updating rule we explained above.
Once the last tournament of the last season ends, the simulation can stop (in \cref{fig:model-draw} the situation of the simulated draw at the end of a generic tournament is shown as an example).
Then, it is possible to extract several outcomes as function of the chosen talent strength $a$, which is the only free control parameter of the model (see \cref{eq:serv}).

\section{3. Simulation Results and Comparison with Data} 

\subsection{3.1 Finding the optimal range of values for the talent strength parameter}
\label{subsec:overall}
 
In what follows, we describe how we tuned the talent strength $a$ in order to find, through an extended simulations campaign, the optimal range of values making our model able to fit real data over $10$ seasons from $2010$ to $2019$.
This allowed us, on one hand, to quantify the role of chance in determining the success of tennis players and, on the other hand, to find the relationship between ranking and talent -- hidden in the real world -- by the unavoidable contribution of random external factors in the athletes' performance. 

\begin{figure}[h]
\centering
\includegraphics[width=0.95\textwidth]{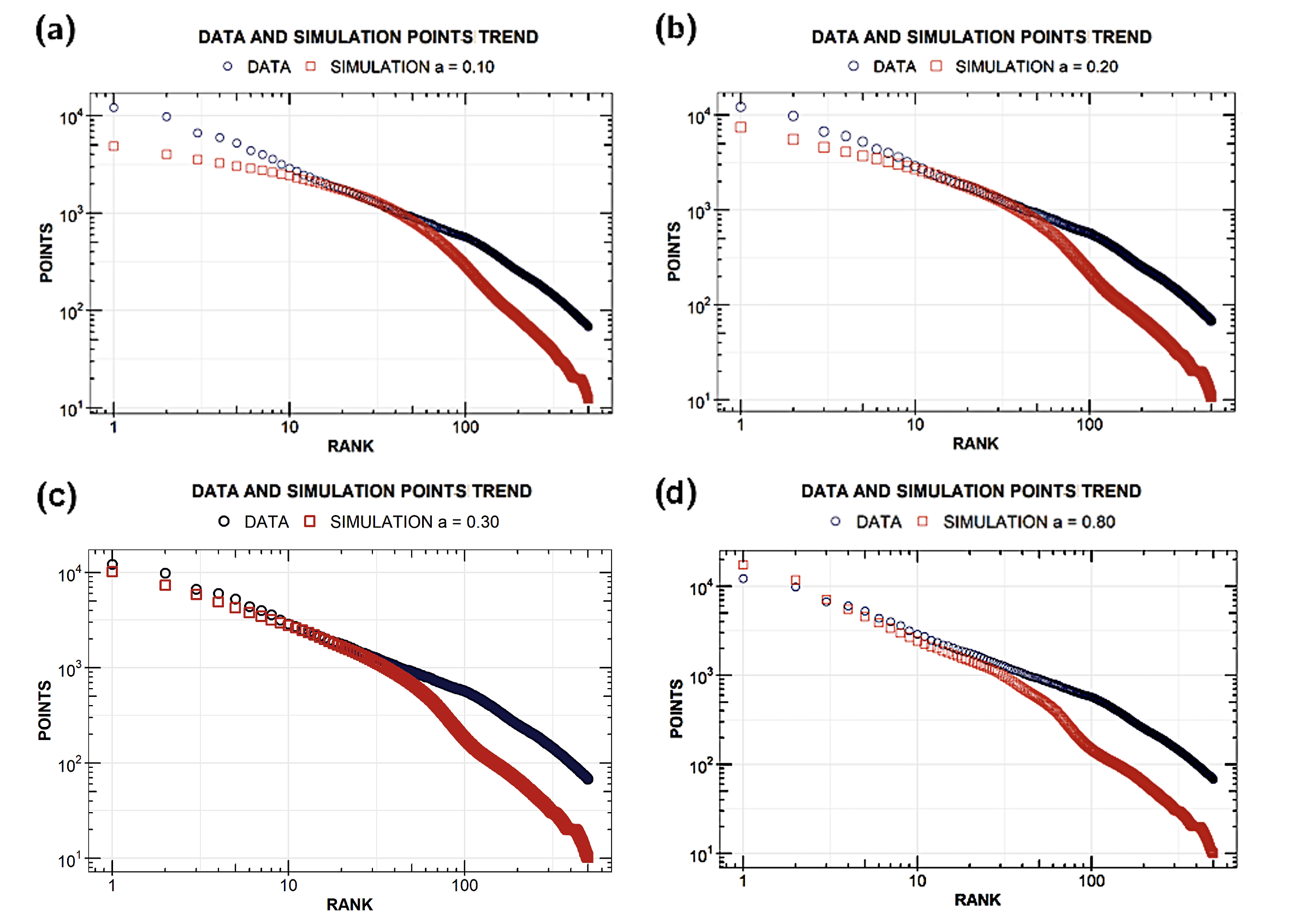}
\caption{Trend of the average total points as a function of ranking placements, in both data and simulation, for the following values of $a$: $a=0.1$ (a), $a=0.2$ (b), $a=0.3$ (c) and $a=0.8$ (d).}
\label{fig:tpoints_datasim}
\end{figure}

At first, we considered the total points cumulated in average during each of the $10$ seasons by the players in the ATP circuit as a function of the ranking placement.
In the four panels of \cref{fig:tpoints_datasim} we compare these data with the simulated ones by plotting the behaviour of total points vs ranking placement for several values of the talent weight $a$, namely $a=0.1$ (a), $a=0.2$ (b), $a=0.3$ (c) and $a=0.8$ (d).
At first sight, the agreement seems quite good only when we consider the top $50$ players, for which the total points scale following a power-law.
This is due to the fact that real ranking positions are also shaped by a lot of minor tournaments not present in our dataset, as explained in Section 2.2.
In particular, focusing on the first $50$ players, the best agreement between data and simulations is obtained for values of talent strength between $a=0.2$ and $a=0.3$, i.e. when talent matters between $20\%$ and $30\%$ with respect to chance.
These values are quite low and, if confirmed, would mean that the impact of random fluctuations on each single point in a tennis competition cannot be absolutely neglected.   

Focusing on panels (b) and (c), it can be noticed that the simulation points for $a=0.2$ seem to better capture the real behaviour of players between the \nth{10} and the \nth{50} positions in the ranking, while the simulation points for $a=0.3$ better describe what is observed for the best ten players in the ranking.
Such a feature could be interpreted as the symptom that athletes in the first ten positions are able to control randomness slightly better than the other ones, thus managing to be more effective in exploiting their talent.   
 
\begin{figure}[h!]
\centering
\includegraphics[width=0.9\textwidth]{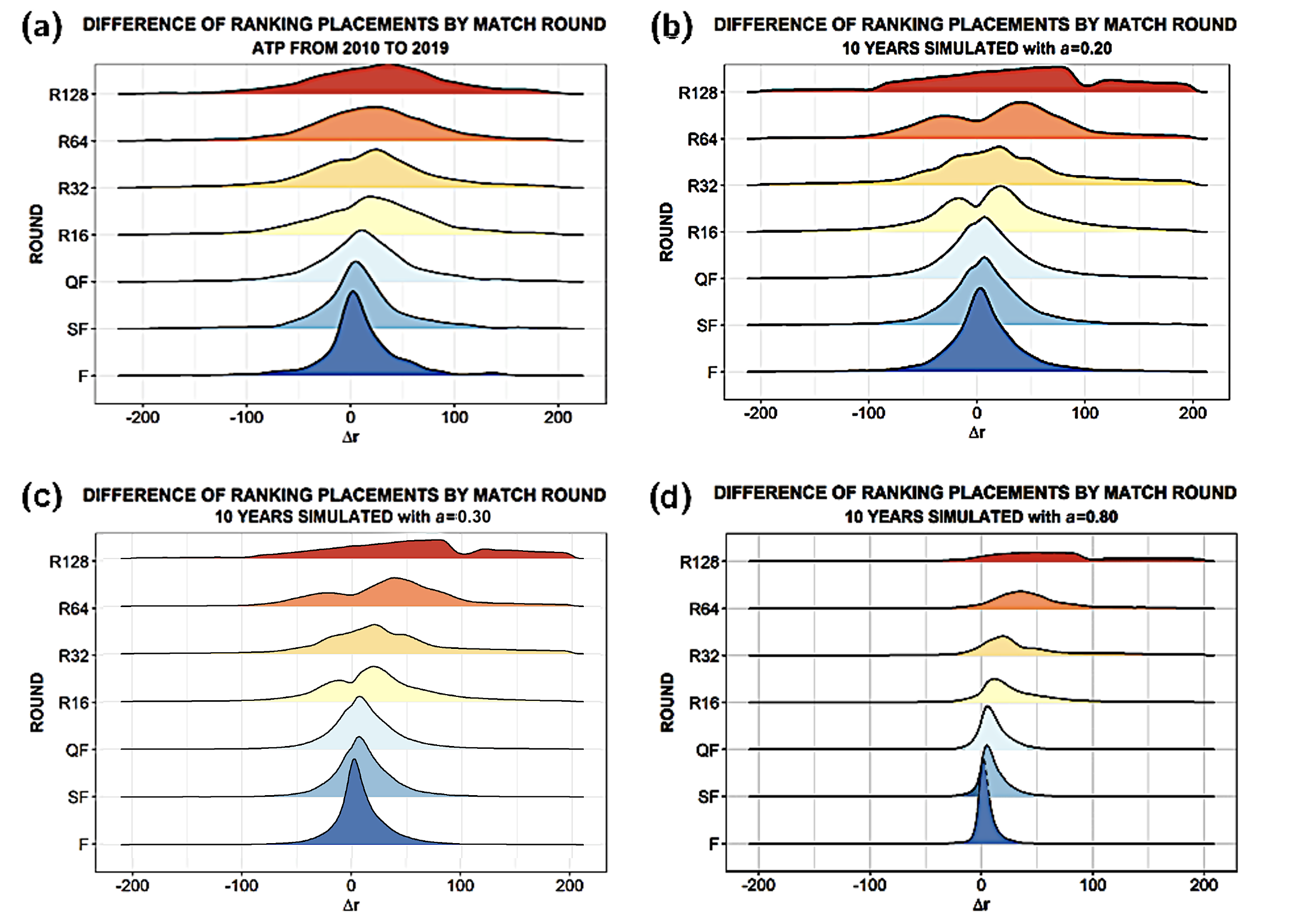}
\caption{Distributions of ranking differences in tennis matches, organized (from top to bottom) by rounds of the main draw.
Panel (a) shows ATP data results, while the other panels reports simulated results obtained for $a=0.2$ (b), $a=0.3$ (c) and $a=0.8$ (d).}
\label{fig:rounds}
\end{figure}

In order to investigate the ability of our model in reproducing all the stages of real ATP competitions, let us now look at the difference between ranking positions of winners and losers, $r_{winner}$ and $r_{loser} $ respectively, in the various rounds of the main draw of a generic tournament.
In detail, we define the difference $\Delta r$ as follows:
\begin{equation}
\Delta r = r_{loser} - r_{winner}
\label{eq:diff_rank}
\end{equation}
and remark that higher values of $r$ mean worse placements in ranking.
In that sense, $\Delta r > 0$ implies that players with a better position in ranking win; on the other hand, $\Delta r < 0$ indicates that players in lower placements win.
An increase in $\vert \Delta r \vert$ indicate players become more distant in the ATP ranking.
Note that $r$ identifies the position in ranking at the time when the match between the two opponents occurred.
In \cref{fig:rounds} we report the distributions of $\Delta r$, calculated round by round for both ATP data (panel a) and simulation runs in correspondence of three increasing values of $a$ (panels b, c and d).
To enhance visualization, we neglect values of  $\vert \Delta r \vert > 200$.

In panel (a) of \cref{fig:rounds} we observe a general tendency of the peaks of the distributions to shift towards $0$ when one goes from the round of $128$ to the final $F$.
This trend is likely due to a reduction in the ranking difference between the players as they reach the last rounds of the tournaments and is quite well captured by our model, in particular when $a=0.2$. 
However, we see a few discrepancies between data and simulations, since in the latter bi-modal distributions emerge for some rounds, while they are less evident in data curves. Those differences might derive from the approximation we made in the number of ATP $250$ tournaments per year and, more generally, in the maximum number of participants per tourney type.
In any case, simulation outputs correctly reproduce the decrease of the variance going from the first rounds (round of $128$, $64$, etc...) to the last ones (quarter-finals, semi-finals, finals).
The model just fails in replicating the distribution of $\Delta r$ at round of $128$, underlining again that it better captures the evolution of the top $50$ players (who are, typically, more likely to reach the final rounds in main tournaments).

\begin{figure}[h!]
\centering
\includegraphics[width=0.98\linewidth]{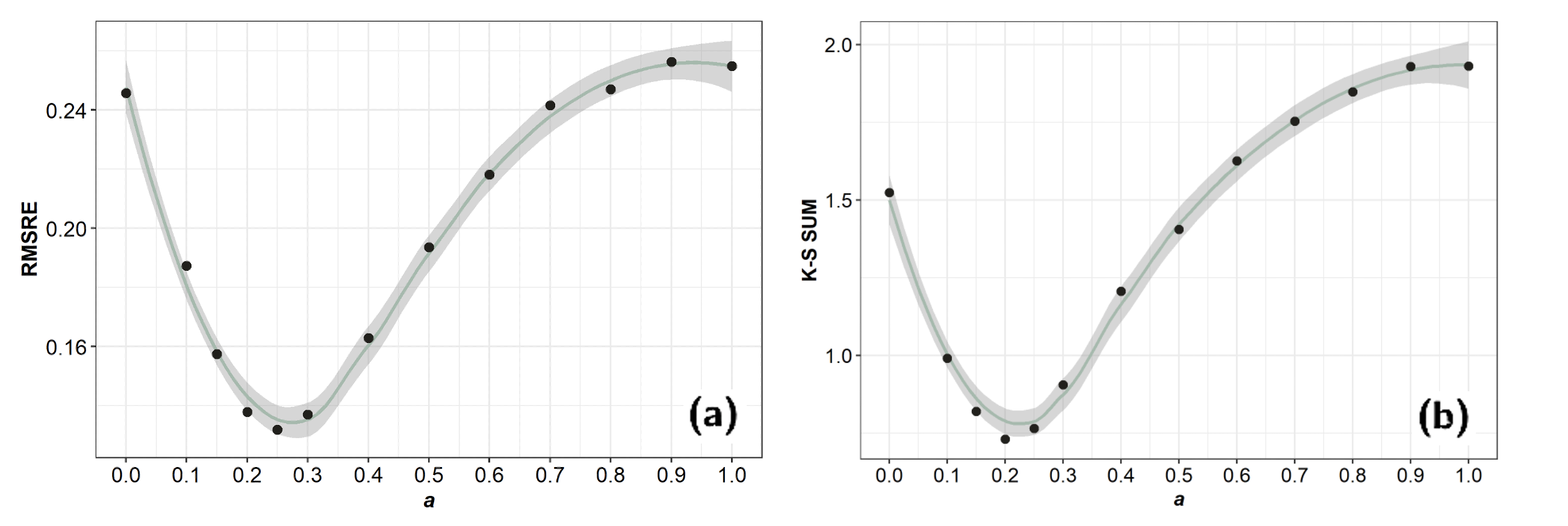}
\caption{Panel (a): The root mean squared relative error (RMSRE) between ATP data and simulations, calculated for the total points vs ranking, is reported as function of the talent strength $a$. Panel (b): The K-S Sum, calculated through a Kolmogorov-Smirnov test applied to the distributions of the ranking differences $\Delta r$ for ATP data and simulations, is reported as function of $a$. In both panels, a green solid line interpolating all the points is also reported together with its confidence interval, in gray shadow.
}
\label{fig:rmspe_errs}
\end{figure}

In \cref{fig:rmspe_errs} we identify, in a more quantitative way, what is the range of values of talent strength which is able to ensure the best fit of ATP data with our simulation results.
In panel (a) the root mean squared relative error \cite{Gken2016IntegratingMA} (RMSRE) between data and simulation for the total ranking points is reported as function of $a$, while in panel (b) the Kolmogorov-Smirnov test \cite{Feller1948} (K-S), performed to compare the distributions of the ranking differences $\Delta r$ in ATP data and simulations, is also reported as function of $a$.
One can easily see that, for both the curves interpolating the RMSRE (a) and the K-S Sum (b) points (represented as solid lines surrounded by their confidence intervals), the minimum falls -- in agreement with our intuition coming from the observation of the previous figures -- within the range $[0.20, 0.30]$ of the talent strength $a$. 

This result, which seems to confirm the fundamental role of chance in determining the outcomes of tennis competitions (at least for male players), is surely quite unexpected for those considering the success an expression of the individual talent of tennis players.
However, \emph{a posteriori}, it can be easily explained by the characteristics of the data we analysed and we based our model on, since they capture only the behaviour of players at the top positions in ATP circuit.
As a matter of fact, in this case a paradoxical mechanism (also conjectured by Mauboussin  \cite{mauboussin2012success}) clearly takes place: the higher the level of competitiveness, the more similar the individual abilities of players, the less determinant talent becomes to the match outcome. Matches between players in the first $50$ placements are often influenced by few, significant turning points.
Our model strongly supports such conclusion, thus suggesting its validity for any other individual sport where tournaments are organized in successive rounds, to be validated in future researches.

\begin{figure}[h]
\centering
\includegraphics[width=0.9\textwidth]{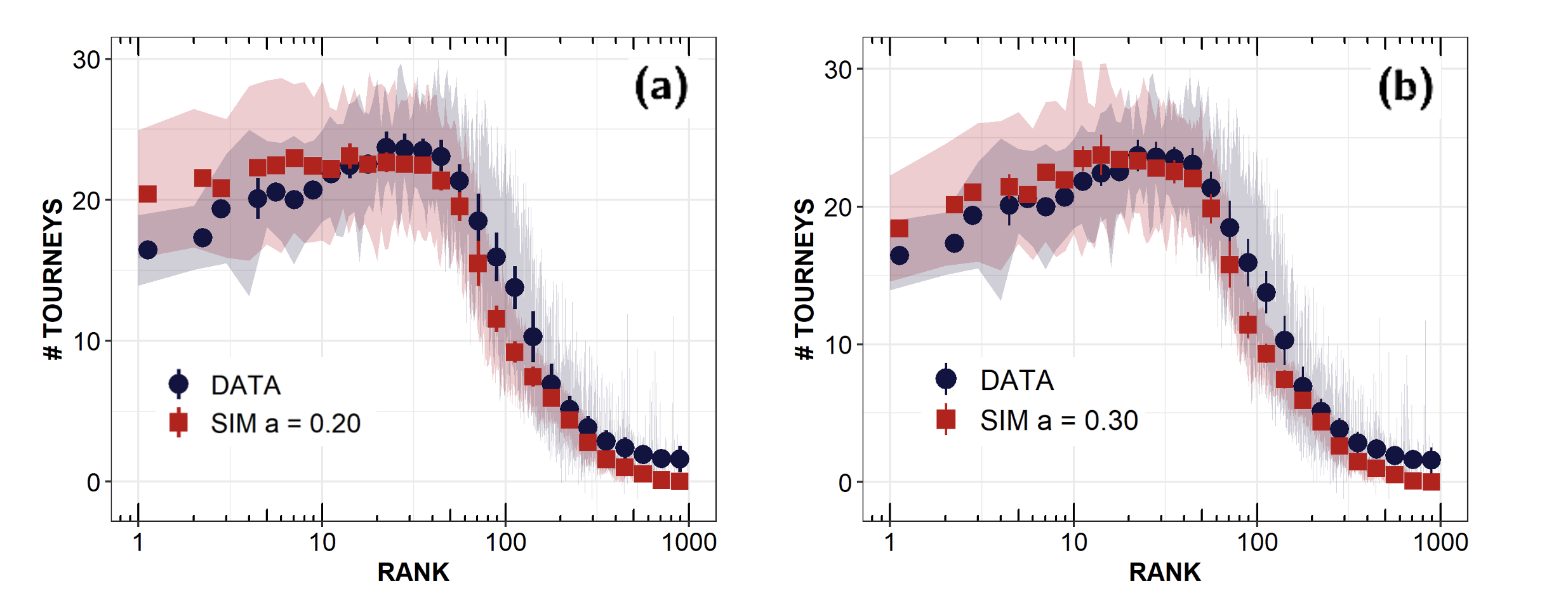}
\caption{Tourney attendance as a function of ATP ranking. Comparison between data (blue dots) and simulations (red squares) with either talent strength $a=0.2$ (a) or $a=0.3$ (b).}
\label{fig:ntour_comps}
\end{figure}

Finally, we checked that, once calibrated with $a=0.2$ or $a=0.3$, our model is able to correctly replicate the real average number of tournaments attended in a season, as a function of players' ranking (see \cref{fig:ntour}).
Results are shown in the two panels of \cref{fig:ntour_comps}, where we see a good agreement between data and simulations, in both the cases within the standard deviation (shaded area in the figures), for the average number of tournaments per ranking placement in a season. The slightly better performance of simulations with $a=0.30$ for the top ten players could be linked with the analogous better agreement of their average total points vs ranking noticed in \cref{fig:tpoints_datasim} (c). 

\subsection{3.2 Inspecting the features of the agents}
\label{subsec:agents}

In our model, we assume agents' talent do not vary during a simulation run.
This is a simplification, and one can argue it neglects the role of training, learning and physical changes in the evolution of a professional career in sports \cite{Sobkowicz2020}.
However, we maintain that a dynamic definition of talent might introduce biases, thus reducing the interpretation of model results. Talent should be intended as a combination of multiple intrinsic characteristics of a person, relevant to achieve results: without loss of generality we assume, quite naturally, that the impact of talent variations is negligible in professional performers at the highest level of practice \cite{Sobkowicz2020}.
We interpret athletes' ability to perform to their highest potential during competitions as talent \cite{Zappala2022}.
\begin{figure}[h]
\centering
\includegraphics[width=\textwidth]{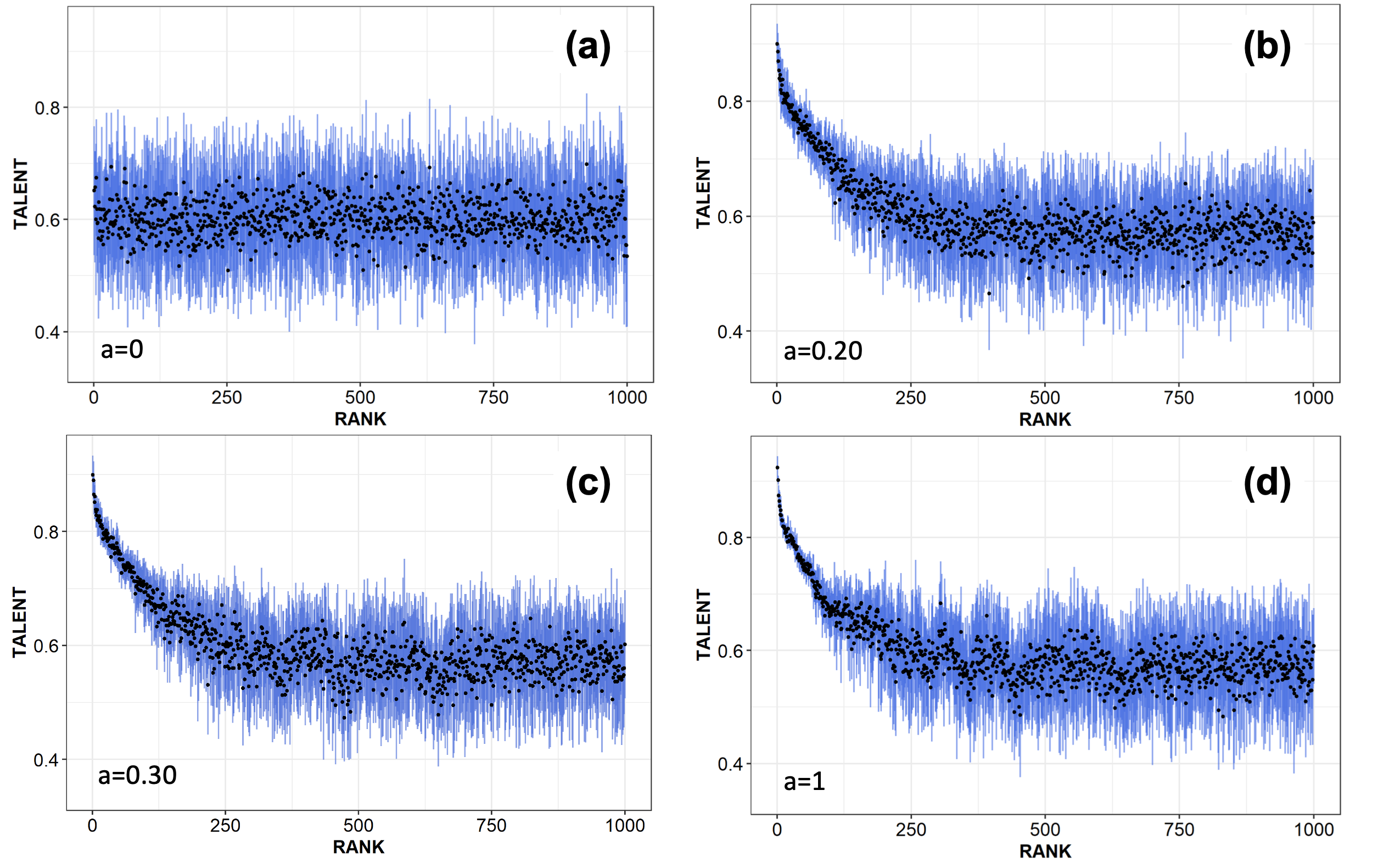}
\caption{Talent as a function of ranking for different values of $a$.
Black dots mark the mean values, while the blue bars represent their standard deviations.}
\label{fig:tals}
\end{figure}

In \cref{fig:tals} we observe the trend of the players' talent, averaged over the 10 seasons, as a function of their ranking position, for four distinct scenarios expressed by the talent strength $a$: the completely random one, $a = 0$ (a); the reference cases $a = 0.20$ (b) and $a=0.30$ (c); the ideal case $a = 1.00$ (d). Panel (a), obviously, displays a complete absence of correlation between ranking and talent, given that, for $a=0$, the latter does not contribute at all to the athletes' performance. On the contrary, as shown in panels (b), (c) and (d), when $a>0$ a strong correlation between talent and ranking becomes evident for the first two-hundred positions; in particular, starting from its maximum value $T\sim0.9$ for the first position in the ranking, a very rapid collapse of the average talent can be observed along the top thirty players, followed by a slower -- almost linear -- decreasing trend which holds until the \nth{250} position; then the correlation disappear and talent starts to fluctuate around the mean talent ($T=0.6$) independently from the ranking, meaning that one can find any athlete with talent $0.5<T<0.7$ in any position between $250$ and $1000$.  
Interestingly, all these three panels appear to be very similar one with the other, apart from the smaller standard deviations of the top fifty players when $a = 1.00$. 
These results, on one hand confirm that players at the bottom of the ranking are not necessarily the less talented but could be moderately gifted athletes who, simply, haven't had yet the chance to show their capabilities. On the other hand, they reinforce the paradoxical mechanism of the influence of talent, which is correlated with ranking almost in the same way either in simulations where it weights less than chance ($a = 0.20$ and $a=0.30$) or in simulations where it represents the main contribution for success, thus revealing that other factors affect the dynamics of ATP ranking.

It is worth noting that \cref{fig:tals} have no equivalent in data, being talent an hidden variable of players.
Moreover, the set of agents is constant during a given simulation run, as opposite to the community of players in our dataset.
Perhaps, that aspect might favour the correlation between the first ranking positions and athletes' talent.
However, such a correlation overall is not as strong as one could expect after the selective processes of tournaments.
This suggests that talent may be neither well preserved nor fairly rewarded in sports like tennis, and assumptions of this kind should be avoided in general.

\subsection{3.3 Focusing on ranking interactions}
\label{subsec:interact}
In individual sports based on direct elimination tournaments, pairwise interactions among the players are crucial, since winning a competition depends on the outcome of single matches, from the point of view of the individuals. 
For this reason, we study the network of matches from 2010 to 2019 of male tennis players in the ATP tour and compare it to the network built by simulated matches of our agent-based model.

\begin{figure}[p]
\centering
\includegraphics[width=.9\textwidth]{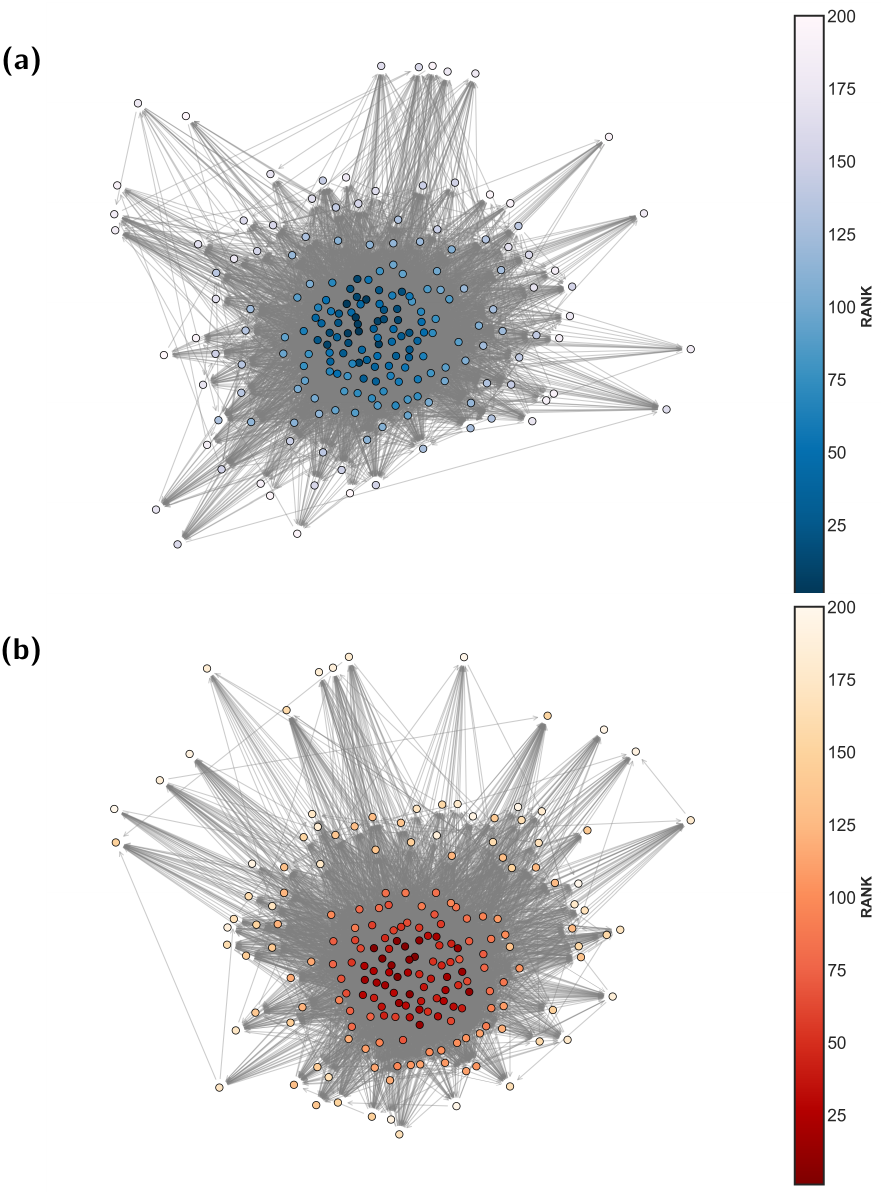}
\caption{Network visualization of the top 200 ranked nodes. Panel (a) shows the data layout, while panel (b) refers to simulations with $a=0.20$. In both  cases a core-periphery structure can be easily identified.}
\label{fig:nets}
\end{figure}

In the previous sections we investigated the dynamics of ranking, considering how tournaments affect it but neglecting the details of the head-to-head results, as a function of ranking placements.
Rather, we focused on the influence of tournament rounds, precisely the main draw ones. In the following we consider a weighted directed network where nodes are positions in ATP ranking, and a link $\left(a, b\right)$ is created if $a$ loses against $b$ in any round of the main draw of the top-tier ATP tour without distinction, thus disregarding qualification stages.
This choice for the structure of the network follows the rationale of \cite{Lai2018}; it sounds reasonable since it highlights the importance of nodes in terms of their ingoing connections, which is common practice in network science literature. 

We remind that the adjacency matrix for a directed network has elements \cite{newman2018networks}: 
\begin{equation}
    A_{ij}=
    \begin{cases}
      1 & \text{if there is an edge from}\ j \quad \text{to}\ i \\
      0 & \text{otherwise}
    \end{cases}
\label{eq:adj_mat}
\end{equation}

When we consider the weights of the links, $A_{ij} = 1$ can be replaced by $A_{ij} = w_{ij}$, where $w_{ij}$ is the number of times $i$ defeats $j$.
Specifically, we normalize $w_{ij}$ taking into account the total number of times $i$ faces $j$:
\begin{equation}
\tilde{w}_{ij} = \frac{w_{ij}}{w_{ij} + w_{ji}}
\label{eq:norm_weights}
\end{equation}

We study two networks: one built from our dataset, with $N = 667$ nodes and $L = 25413$ edges, with the condition that the ATP ranking of nodes must be less than or equal to 1000; the other graph results from simulation runs with the reference value of the talent strength $a = 0.20$, sampling links until we obtain the same value of $L$, resulting in a network with $N_{sim} = 986$.

In \cref{fig:nets} we first visualize \cite{kamada1989algorithm} the two networks for a restricted set of nodes, the first 200 in ranking, for both real data (a) and simulated ones with $a = 0.20$ (b). By indicating the nodes' rank with a colour scale, in both cases one can spot the presence of a denser core of top-ranked nodes surrounded by a periphery of less connected nodes indicating lower positions in the ranking. This common features is not surprising, since top-ranked players more likely end to play one against each other in the final rounds of the main competitions. However, it confirms the ability of our model to reproduce, also from a topological perspective, the behaviour of the ATP dataset when a talent strength $a=0.20$ is assumed.  

To further reinforce this last statement, we compare the in-degree and out-degree distributions of both the networks.
The in-degree is the number of ingoing links connected to a node and the out-degree is the number of outgoing links.
Bearing in mind the adjacency matrix \cref{eq:adj_mat}, we can write:

\begin{equation}
k_{i}^{in} = \sum_{j = 1}^{N} A_{ij}, \quad k_{j}^{out} = \sum_{i = 1}^{N} A_{ij}
\end{equation}

\begin{figure}[h!]
\centering
\includegraphics[width=0.8\textwidth]{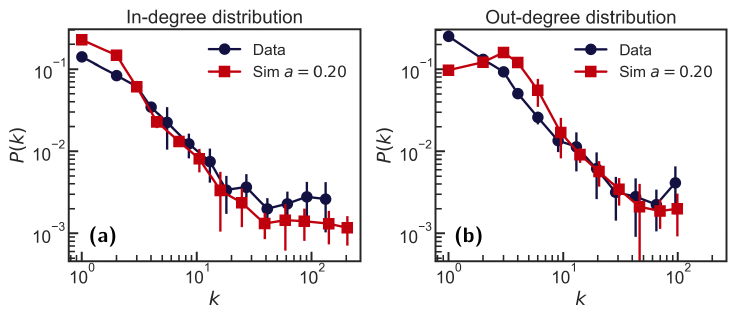}
\caption{In-degree (a) and out-degree (b)  distributions for the networks of both real (circles) and simulated (square) matches.}
\label{fig:degrees}
\end{figure}

\begin{figure}[h!]
\centering
\includegraphics[width=0.8\textwidth]{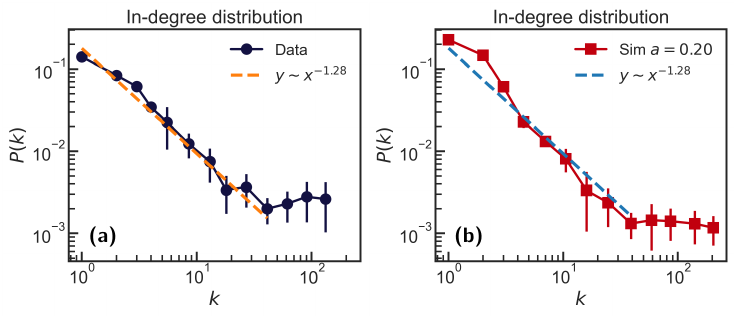}
\caption{Power law fit of the in-degree distribution for both real ATP data (a) and simulations (b).}
\label{fig:indeg_fit}
\end{figure}

In \cref{fig:degrees} we observe a very good agreement between data and simulation for the in-degree distribution (a), which scales as a power law if we exclude the tail (see \cref{fig:indeg_fit}); instead, for the out-degree distributions (b), the two trends differ for low values of $k$. 

To show the relationship between the importance of nodes in the network and their ranking positions in the ATP, we test two centrality measures.
The former is the in-degree centrality:
\begin{equation}
c_{i}^{in} = \frac{1}{N - 1} \sum_{j = 1}^{N} A_{ij}
\label{eq:indeg}
\end{equation}
which in this case captures how central certain nodes, representing the ranking placements of players with many wins, become in the topology of the network.

\begin{figure}[t]
\centering
\includegraphics[width=0.8\textwidth]{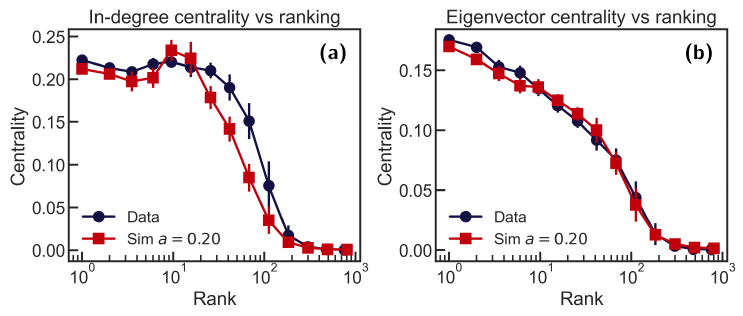}
\caption{In-degree centrality (a) and eigenvector centrality (b) of the two networks of matches are reported as a function of ranking placements. The agreement between real ATP data and our model simulations for $a=0.20$ is in both  cases very good.}
\label{fig:cents}
\end{figure}

Panel (a) of \cref{fig:cents} shows the in-degree centrality as a function of players' ranking placements in both data and simulations.
We observe an overall concordance in trend, although simulation curve slightly underestimates in-degree centrality values.
In more detail, the top fifty positions have approximately the same centrality, meaning that they have a comparable number of wins, on average.

A natural extension of the in-degree centrality is the eigenvector centrality, which awards nodes not only because they have many connections, but also because they are connected to other nodes which are themselves important \cite{newman2018networks}.
In the network of tennis matches, it underlines the fact that winning many times can be relevant but not sufficient: a victory turns out to be crucial also when the winner player defeats an opponent who has  already a lots of wins at his credit.

The eigenvector centrality $x_{i}$ for a node $i$ in a directed network is proportional to the centralities of the nodes that point to $i$:
\begin{equation}
x_{i} = \kappa_{1}^{-1} \sum_{j} A_{ij} x_{j}
\label{eq:eigen}
\end{equation}
where $\kappa_{1}$ is the largest eigenvalue of the adjacency matrix \cite{newman2018networks}.
We specify that \cref{eq:eigen} holds for weighted networks.
Using such a definition, given that we consider ranking placements as nodes, we expect a monotonic decreasing relationship between ranking and eigenvector centrality.
That is exactly what we observe in panel (b) of \cref{fig:cents}; also, eigenvector centrality curve is correctly predicted by the model simulations.
It is worth noting that the decreasing trend of eigenvector centrality is evident only for the top 100 players, whereas it is less significant for the other placements.

Interestingly, the comparison between the two analysed centrality measures captures some features of our specific context, which is men's professional tennis.
On one hand, in-degree centrality, which can be interpreted as the importance of how much players win, put all the top ranked positions on the same level; in other words, one could not distinguish between them, based on in-degree only.
On the other hand, eigenvector centrality adds the not negligible weight of defeating high centrality players: by taking this into account, a scale emerges, thus making the top placements distinguishable from each other.
Interestingly, both those behaviours are correctly reproduced by our model with $a=0.20$: despite the limited weight of talent, a selective mechanism still arises.

\begin{figure}[h]
\centering
\includegraphics[width=0.95\textwidth]{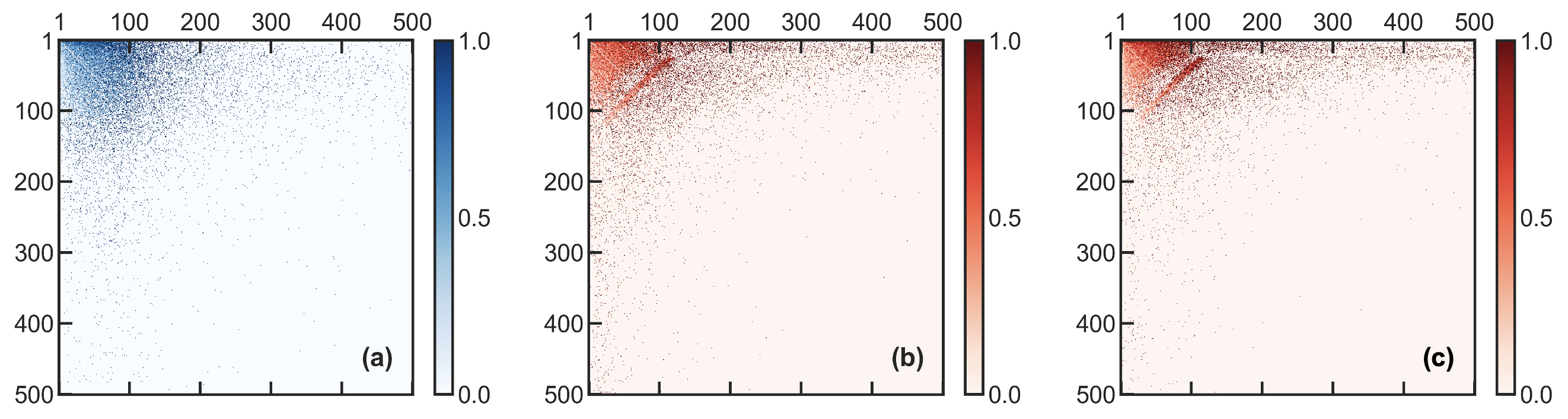}
\caption{Adjacency matrix of 500 nodes for real ATP data (a) and simulations for $a=0.2$ (b) and $a=0.30$ (c).
Colorbars highlight the different weights of the elements.}
\label{fig:adj_mats}
\end{figure}

To better understand the interplay between the ranking and the structure of the network of matches, we inspect their adjacency matrices representing the weight of each element with a color scale.
\cref{fig:adj_mats} shows the two weighted matrices for real data (a) and simulations with $a=0.20$ (b): in line with the visualization of \cref{fig:nets}, they appear both sparse and asymmetrical. In panel (c) we also add the adjacency matrix for simulations with the other reference value chosen for comparison, $a=0.30$, in order to appreciate possible differences. In this case, at a first sight, these differences seem negligible. 

\begin{figure}[h]
\centering
\includegraphics[width=0.7\textwidth]{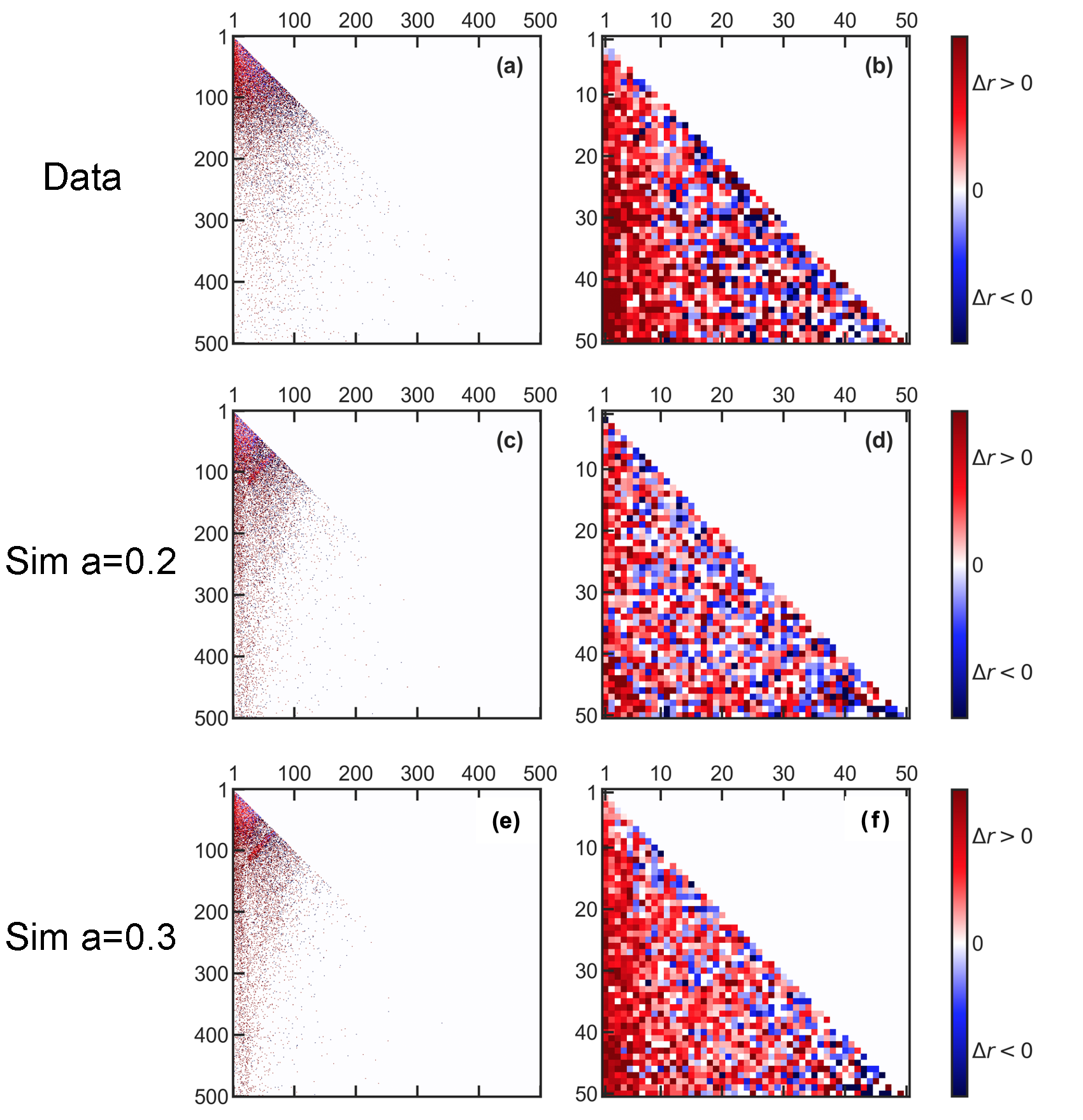}
\caption{Triangular matrix of the excess of wins given the different ranking of the players.
Upper panels (a-b) refer to real ATP data matches; middle panels (c-d) refer  to simulations with talent strength $a=0.20$; bottom panels (e-f) refer  to simulations with talent strength $a=0.30$.
Left panels (a-c-e) show 500 nodes, right panels (b-d-f) are an inset of the first 50 nodes.}
\label{fig:tri_mats}
\end{figure}

To further inspect these three networks, let us consider a triangular matrix whose elements are $\widehat{A}_{ij} = A_{ij} - A_{ji}$.
In this way, we evaluate whether players in ranking positions $i$ or $j$ won more times when they faced each other, since the elements $A_{ij}$ are the fraction of wins $\tilde{w}_{ij}$ of the node $i$ and, vice-versa, $A_{ji}$ is the ratio of victories $\tilde{w}_{ji}$ of the node $j$ (see \cref{eq:norm_weights}).
Results are shown in the three panels of \cref{fig:tri_mats}. Provided that $\widehat{A}_{ij} = 0 \quad \forall i < j$, this representation gives us insights about the ranking difference $\Delta r$ between $i$ and $j$, as underlined in the color-bars of the heat-maps: following the definition given in \cref{eq:diff_rank}, red elements of the matrix indicate better ranked players win more, while values in blue mean worse ranked players win more.
The case of $A_{ij} = A_{ji}$, which would represent a balanced number of victories between $i$ and $j$, cannot be distinguished from the absence of matches $i \leftrightarrow j$ with this formalism; however, this occurs only in $0.5 \%$ of the dataset.

Comparing panels (a), (c) and (e) in \cref{fig:tri_mats}, one can immediately see that both data and simulations show a higher density for the first 100 nodes. However, while for the real matches top players have a low probability to face athletes beyond the \nth{200} placement, for the simulated matches (with both $a=0.2$ and $a=0.3$) top agents seem more likely to face opponents in the bottom part of the ranking.
Focusing on panels (b), (d) and (f), we see that our model correctly captures the unpredictability of a match outcome as we approach the diagonal of the matrix, i.e. when ranking positions are closer.
In fact, in addition to the general color mixing with no patterns characterizing all the networks, the common presence of many blue areas indicates higher possibilities for better ranked players of being defeated. 
On the other hand, the simulated network with $a=0.2$ (d) does not exactly reproduce the pattern off-diagonal: in  fact, the dominance of the top 10 placements, evident as dark red area at the bottom of the first 10 columns in panels (b), of real data, and (f), of the simulated network with $a=0.3$, is less pronounced in panel (d). This feature seems in line with the results shown in \cref{fig:tpoints_datasim}, where the simulated points for the first ten position in the ranking were better captured by a talent strength $a=0.3$ rather than $a=0.2$, thus confirming a possible greater role of talent in the performance of the top players of the ranking. 

\section{4. Discussion and conclusions}
\label{sec:discussion}

In this work, we studied the impact of luck in individual sports.
Specifically, we investigated the role of talent and chance in tennis, focusing on the ATP circuit and quantifying the influence of unpredictable factors, from the level of single points in a match to the macroscopic evolution of the whole ranking. 

We built an agent-based model and calibrated it on data from international rankings and tournaments.
This model allowed us to estimate the relative weight of chance compared to talent in ATP tournaments.
The starting hypothesis was that small random fluctuations at the level of a single point, along with the structural ``bias'' characterising tennis and resulting in an asymmetrical probability of winning points when at service or receiving, could become relevant, especially when the level of competitiveness is very high and selective mechanisms of tournaments bring athletes with a comparable, very high, talent to confront each other. 

Due to the availability, the amount and the organization of data, we limited our consideration to the main tourneys of the ATP tour (Grand Slam, Masters 1000, ATP 500 and 250).
We collected matches and rankings of the first thousand male athletes in the ATP circuit from 2010 to 2019.
Through a comparative analysis, where the weight $a$ of talent was the only free parameter, we found that our model can reproduce relevant features of that community of athletes -- e.g. the distributions of points as a function of ranking and the distributions of ranking differences in distinct match rounds -- for the values of $a$ included in the interval $[0.20,0.30]$.
This result would imply that random events weigh consistently in determining tennis matches, on average around $70\%-80\%$ for men's professional players.

Under this assumption, we also showed that our agent-based model captures the main elements of the directed network built from the match results between ranking positions.
Specifically, the network obtained from our numerical simulations shows centrality trends comparable with the network of the data for both the in-degree centrality and the eigenvector centrality.
The former reflects the importance of how much players win.
The latter underlines the influence of defeating high-centrality players.
Finally, simulated networks allowed us to explore the dominance of certain ranking positions over others.

Although our model gives valuable insights into the evolution of tennis careers, it also has some evident limitations.
Firstly, we simulate a community of 1000 players/agents that do not change in time along the 10 considered seasons, while the community of real athletes is an open system, where someone continually withdraws, and someone else enters the circuit.     
Furthermore, ATP rankings are not limited to the main tournaments of the ATP tour in reality; for this reason, the model cannot reproduce the overall ranking dynamics.
This explains why some results apply to a limited set of players.
Secondly, the talent of each simulated athlete is also fixed in time, thus representing the maximum level of ability he can reach throughout his entire career while considering both individual skills and efforts.
Of course, this is a strong approximation since any player can improve their performance with the help of hard training and cumulated experience over ten years.
Moreover, in our model, the influence of chance is limited to the outcome of each single point and to the asymmetry of service and return, which means that an external source of noise, independent from tennis rules, is absent.
Such an external source could be represented, for example, by the random occurrence of injuries or accidents that can condition, interrupt or even defeat the promising career of many brilliant players.

Despite such limitations of the model, we believe that our study could provide several valuable perspectives.  
In fact, by fixing the relative weight of talent between $20\%-30\%$, it gives further support to the existence of a paradoxical mechanism (already claimed in \cite{mauboussin2012success}): i.e. the higher the level of competitiveness, the closer the talent of athletes involved, the more determinant random fluctuations become to win a match and to develop a successful career.
The evidence of a ``rich-get-richer'' effect should question the current awarding methods, rooted in the dominating meritocratic paradigm, and discourage the application of an unfair ``winner-takes-all'' logic. In particular, we consider the configuration of rewards and their distribution to be worth a specific investigation.
Divergences in monetary rewards -- and consequent non-monetary rents -- appear to depict a profoundly different picture than the one suggested by the talent comparison.
This might have a role in excessively reducing the remuneration of the loser in finals or semifinals when compared to the difference in talent with the opponent.
Consequently, athletes with similar skills end up with very different outcomes and wealth profiles, simply because of small fluctuations, at the level of single points of a match.

Having observed similar results in other sports based on tournaments \cite{Zappala2022}, the general methods of awarding should be seriously questioned, both in terms of ranking points and prizes, which typically follow an exponentially decreasing trend, from the first classified to the last one.
Such a configuration is probably more desirable to attract a more sensationalistic environment for each challenging match, but it remains harmful to the sports discipline and for the athletes's career.
The assumption of a one-to-one correspondence between players' talent and their performance in competitions is unlikely to exist, as randomness can significantly influence any athlete's performance and the match itself.
Our future studies will focus on the need for a revision in the current awarding rule, with the aim of enabling a more unbiased and fair  rewarding of talent \cite{Zappala2022}.

\bibliography{library}

\begin{thebibliography}{10}
\urlstyle{rm}
\expandafter\ifx\csname url\endcsname\relax
  \def\url#1{\texttt{#1}}\fi
\expandafter\ifx\csname urlprefix\endcsname\relax\def\urlprefix{URL }\fi
\expandafter\ifx\csname doiprefix\endcsname\relax\def\doiprefix{DOI: }\fi
\providecommand{\bibinfo}[2]{#2}
\providecommand{\eprint}[2][]{\url{#2}}

\bibitem{Sinatra2016}
\bibinfo{author}{Sinatra, R.}, \bibinfo{author}{Wang, D.},
  \bibinfo{author}{Deville, P.}, \bibinfo{author}{Song, C.} \&
  \bibinfo{author}{Barab{\'{a}}si, A.~L.}
\newblock \bibinfo{journal}{\bibinfo{title}{{Quantifying the evolution of
  individual scientific impact}}}.
\newblock {\emph{\JournalTitle{Science}}} \textbf{\bibinfo{volume}{354}},
  \doiprefix\url{10.1126/science.aaf5239} (\bibinfo{year}{2016}).

\bibitem{Fraiberger2018}
\bibinfo{author}{Fraiberger, S.~P.}, \bibinfo{author}{Sinatra, R.},
  \bibinfo{author}{Resch, M.}, \bibinfo{author}{Riedl, C.} \&
  \bibinfo{author}{Barab{\'{a}}si, A.~L.}
\newblock \bibinfo{journal}{\bibinfo{title}{{Quantifying reputation and success
  in art}}}.
\newblock {\emph{\JournalTitle{Science}}} \textbf{\bibinfo{volume}{362}},
  \bibinfo{pages}{825--829}, \doiprefix\url{10.1126/science.aau7224}
  (\bibinfo{year}{2018}).

\bibitem{Williams2019}
\bibinfo{author}{Williams, O.~E.}, \bibinfo{author}{Lacasa, L.} \&
  \bibinfo{author}{Latora, V.}
\newblock \bibinfo{journal}{\bibinfo{title}{Quantifying and predicting success
  in show business}}.
\newblock {\emph{\JournalTitle{Nature communications}}}
  \textbf{\bibinfo{volume}{10}}, \bibinfo{pages}{1--8} (\bibinfo{year}{2019}).

\bibitem{Frank2016}
\bibinfo{author}{Frank, R.~H.}
\newblock \emph{\bibinfo{title}{{Success and luck: Good Fortune and the Myth of
  Meritocracy}}} (\bibinfo{year}{2016}).

\bibitem{mauboussin2012success}
\bibinfo{author}{Mauboussin, M.~J.}
\newblock \emph{\bibinfo{title}{The success equation: Untangling skill and luck
  in business, sports, and investing}} (\bibinfo{publisher}{Harvard Business
  Review Press}, \bibinfo{year}{2012}).

\bibitem{Pluchino2018}
\bibinfo{author}{Pluchino, A.}, \bibinfo{author}{Biondo, A.~E.} \&
  \bibinfo{author}{Rapisarda, A.}
\newblock \bibinfo{journal}{\bibinfo{title}{{Talent versus luck: The role of
  randomness in success and failure}}}.
\newblock {\emph{\JournalTitle{Advances in Complex Systems}}}
  \textbf{\bibinfo{volume}{21}}, \bibinfo{pages}{1--31},
  \doiprefix\url{10.1142/S0219525918500145} (\bibinfo{year}{2018}).
\newblock \eprint{1802.07068}.

\bibitem{Pluchino2018a}
\bibinfo{author}{Pluchino, A.} \emph{et~al.}
\newblock \bibinfo{journal}{\bibinfo{title}{{Exploring the role of
  interdisciplinarity in physics: Success, talent and luck}}}.
\newblock {\emph{\JournalTitle{PLoS ONE}}} \textbf{\bibinfo{volume}{14}},
  \bibinfo{pages}{1--15}, \doiprefix\url{10.1371/journal.pone.0218793}
  (\bibinfo{year}{2018}).
\newblock \eprint{1901.03607}.

\bibitem{Lago2010}
\bibinfo{author}{Lago-Ballesteros, J.} \& \bibinfo{author}{Lago-Pe{\~n}as, C.}
\newblock \bibinfo{journal}{\bibinfo{title}{Performance in team sports:
  Identifying the keys to success in soccer}}.
\newblock {\emph{\JournalTitle{Journal of Human kinetics}}}
  \textbf{\bibinfo{volume}{25}}, \bibinfo{pages}{85--91}
  (\bibinfo{year}{2010}).

\bibitem{Petersen2008}
\bibinfo{author}{Petersen, A.~M.}, \bibinfo{author}{Jung, W.~S.} \&
  \bibinfo{author}{{Eugene Stanley}, H.}
\newblock \bibinfo{journal}{\bibinfo{title}{{On the distribution of career
  longevity and the evolution of home-run prowess in professional baseball}}}.
\newblock {\emph{\JournalTitle{Epl}}} \textbf{\bibinfo{volume}{83}},
  \doiprefix\url{10.1209/0295-5075/83/50010} (\bibinfo{year}{2008}).
\newblock \eprint{0804.0061}.

\bibitem{Yucesoy2016}
\bibinfo{author}{Yucesoy, B.} \& \bibinfo{author}{Barab{\'{a}}si, A.~L.}
\newblock \bibinfo{journal}{\bibinfo{title}{{Untangling performance from
  success}}}.
\newblock {\emph{\JournalTitle{EPJ Data Science}}}
  \textbf{\bibinfo{volume}{5}}, \doiprefix\url{10.1140/epjds/s13688-016-0079-z}
  (\bibinfo{year}{2016}).
\newblock \eprint{1512.00894}.

\bibitem{Sobkowicz2020}
\bibinfo{author}{Sobkowicz, P.}, \bibinfo{author}{Frank, R.~H.},
  \bibinfo{author}{Biondo, A.~E.}, \bibinfo{author}{Pluchino, A.} \&
  \bibinfo{author}{Rapisarda, A.}
\newblock \bibinfo{journal}{\bibinfo{title}{{Inequalities, chance and success
  in sport competitions: Simulations vs empirical data}}}.
\newblock {\emph{\JournalTitle{Physica A: Statistical Mechanics and its
  Applications}}} \textbf{\bibinfo{volume}{557}}, \bibinfo{pages}{124899},
  \doiprefix\url{10.1016/j.physa.2020.124899} (\bibinfo{year}{2020}).

\bibitem{Zappala2022}
\bibinfo{author}{Zappalà, C.}, \bibinfo{author}{Pluchino, A.},
  \bibinfo{author}{Rapisarda, A.}, \bibinfo{author}{Biondo, A.~E.} \&
  \bibinfo{author}{Sobkowicz, P.}
\newblock \bibinfo{journal}{\bibinfo{title}{On the role of chance in fencing
  tournaments: An agent-based approach}}.
\newblock {\emph{\JournalTitle{PLOS ONE}}} \textbf{\bibinfo{volume}{17}},
  \bibinfo{pages}{1--17}, \doiprefix\url{10.1371/journal.pone.0267541}
  (\bibinfo{year}{2022}).

\bibitem{McGarry1997}
\bibinfo{author}{McGarry, T.} \& \bibinfo{author}{Schutz, R.~W.}
\newblock \bibinfo{journal}{\bibinfo{title}{Efficacy of traditional sport
  tournament structures}}.
\newblock {\emph{\JournalTitle{The Journal of the Operational Research
  Society}}} \textbf{\bibinfo{volume}{48}}, \bibinfo{pages}{65--74}
  (\bibinfo{year}{1997}).

\bibitem{Stewart1983}
\bibinfo{author}{Stewart, J.}
\newblock \bibinfo{journal}{\bibinfo{title}{{The Distribution of Talent}}}.
\newblock {\emph{\JournalTitle{Marilyn Zurmuehlin Working Papers in Art
  Education}}} \textbf{\bibinfo{volume}{2}}, \bibinfo{pages}{21--22},
  \doiprefix\url{10.17077/2326-7070.1034} (\bibinfo{year}{1983}).

\bibitem{Buchanan2002}
\bibinfo{author}{Buchanan, M.}
\newblock \emph{\bibinfo{title}{Ubiquity: Why catastrophes happen}}
  (\bibinfo{publisher}{Crown}, \bibinfo{year}{2002}).

\bibitem{tennisrules}
\bibinfo{title}{Tennis rules}.
\newblock
  \bibinfo{howpublished}{\url{https://www.rulesofsport.com/sports/tennis.html}}.

\bibitem{Pluchino2020}
\bibinfo{author}{Challet, D.}, \bibinfo{author}{Pluchino, A.},
  \bibinfo{author}{Biondo, A.} \& \bibinfo{author}{Rapisarda, A.}
\newblock \bibinfo{journal}{\bibinfo{title}{{The origin of extreme wealth
  inequality in the talent vs Luck model }}}.
\newblock {\emph{\JournalTitle{Advances in Complex Systems}}}
  \textbf{\bibinfo{volume}{23}}, \bibinfo{pages}{2050004},
  \doiprefix\url{10.1142/S0219525920500046} (\bibinfo{year}{2020}).

\bibitem{ATP}
\bibinfo{title}{Official site of men's professional tennis| atp tour}.
\newblock \bibinfo{howpublished}{\url{https://www.atptour.com/}}.

\bibitem{ATPdatahub}
\bibinfo{title}{Atp world tour tennis data}.
\newblock
  \bibinfo{howpublished}{\url{https://datahub.io/sports-data/atp-world-tour-tennis-data}}.

\bibitem{ATPgit}
\bibinfo{title}{Tennis data repository}.
\newblock
  \bibinfo{howpublished}{\url{https://github.com/JeffSackmann/tennis_atp}}.

\bibitem{NetLogo}
\bibinfo{author}{Wilensky, U.}
\newblock \bibinfo{title}{Netlogo}.
\newblock \bibinfo{type}{http://ccl.northwestern.edu/netlogo/},
  \bibinfo{institution}{Center for Connected Learning and Computer-Based
  Modeling}, \bibinfo{address}{Northwestern University, Evanston, IL}
  (\bibinfo{year}{1999}).

\bibitem{Shine2003}
\bibinfo{author}{Shine, O.}
\newblock \emph{\bibinfo{title}{The Language of Tennis}}
  (\bibinfo{publisher}{Carcanet Press}, \bibinfo{address}{Manchester},
  \bibinfo{year}{2003}).

\bibitem{Hedges1978}
\bibinfo{author}{Hedges, M.}
\newblock \emph{\bibinfo{title}{The Concise Dictionary of Tennis}}
  (\bibinfo{publisher}{Mayflower Books}, \bibinfo{address}{New York},
  \bibinfo{year}{1978}).

\bibitem{Gken2016IntegratingMA}
\bibinfo{author}{G{\"o}çken, M.}, \bibinfo{author}{{\"O}zçalici, M.},
  \bibinfo{author}{Boru, A.} \& \bibinfo{author}{Dosdoğru, A.~T.}
\newblock \bibinfo{journal}{\bibinfo{title}{Integrating metaheuristics and
  artificial neural networks for improved stock price prediction}}.
\newblock {\emph{\JournalTitle{Expert Syst. Appl.}}}
  \textbf{\bibinfo{volume}{44}}, \bibinfo{pages}{320--331}
  (\bibinfo{year}{2016}).

\bibitem{Feller1948}
\bibinfo{author}{Feller, W.}
\newblock \bibinfo{journal}{\bibinfo{title}{{On the Kolmogorov-Smirnov Limit
  Theorems for Empirical Distributions}}}.
\newblock {\emph{\JournalTitle{The Annals of Mathematical Statistics}}}
  \textbf{\bibinfo{volume}{19}}, \bibinfo{pages}{177 -- 189},
  \doiprefix\url{10.1214/aoms/1177730243} (\bibinfo{year}{1948}).

\bibitem{Lai2018}
\bibinfo{author}{Lai, M.}, \bibinfo{author}{Meo, R.},
  \bibinfo{author}{Schifanella, R.} \& \bibinfo{author}{Sulis, E.}
\newblock \bibinfo{journal}{\bibinfo{title}{{The role of the network of matches
  on predicting success in table tennis}}}.
\newblock {\emph{\JournalTitle{Journal of Sports Sciences}}}
  \textbf{\bibinfo{volume}{36}}, \bibinfo{pages}{2691--2698},
  \doiprefix\url{10.1080/02640414.2018.1482813} (\bibinfo{year}{2018}).

\bibitem{newman2018networks}
\bibinfo{author}{Newman, M.}
\newblock \emph{\bibinfo{title}{Networks}} (\bibinfo{publisher}{Oxford
  university press}, \bibinfo{year}{2018}).

\bibitem{kamada1989algorithm}
\bibinfo{author}{Kamada, T.}, \bibinfo{author}{Kawai, S.} \emph{et~al.}
\newblock \bibinfo{journal}{\bibinfo{title}{An algorithm for drawing general
  undirected graphs}}.
\newblock {\emph{\JournalTitle{Information processing letters}}}
  \textbf{\bibinfo{volume}{31}}, \bibinfo{pages}{7--15} (\bibinfo{year}{1989}).

\end{thebibliography}

\section*{Acknowledgements}

C.Z. would like to thank Luca Gallo for the inspiring discussions and valuable comments and suggestions.
The authors acknowledge the financial support of the project PRIN 2017WZFTZP ``Stochastic Forecasting in Complex Systems'' and also the project MOSCOVID of Catania University.

\section*{Author contributions statement}

A.P. conceived the model, C.Z. analysed the data and made the numerical simulations. A.P., A.R. and A.E.B. supervised the study. All authors wrote and reviewed the manuscript. 

\section*{Data availability statement}

Data analysed in this work are available online \cite{ATPdatahub,ATPgit,ATP}.

\section*{Additional information}

\textbf{Competing interests} The authors declare no competing interests.

\end{document}